\documentclass[onecolumn]{IEEEtran}    
\IEEEoverridecommandlockouts
\usepackage{setspace}
\usepackage[T1]{fontenc}
\usepackage{amssymb, mathtools, eucal}
\usepackage{amsthm}
\usepackage{subcaption}
\usepackage{rotating}

\theoremstyle{definition}
\newtheorem{definition}{Definition}
\newtheorem{theorem}{Theorem}
\newtheorem{proposition}[theorem]{Proposition}
\newtheorem{lemma}[theorem]{Lemma}
\newtheorem{corollary}[theorem]{Corollary}
\newtheorem{remark}{Remark}

\newtheorem{problem}{Problem}

\usepackage{graphicx}
\usepackage[dvipsnames]{xcolor}
\usepackage{color}
\usepackage{tikz}
\usepackage{tcolorbox}

\usepackage{multirow}  % for commutative diagrams
\usetikzlibrary{calc}
\usepackage{arydshln}
\usepackage{multicol}
\usepackage{enumerate}

\usepackage{cite}
\usepackage{url}
\usepackage{hyperref}
\newcommand\myshade{70}
\hypersetup{
  linkcolor  = red!\myshade!black,
  citecolor  = blue!\myshade!black,
  urlcolor   = blue!\myshade!black,
  colorlinks = true,
}

\usepackage{algorithm}
\usepackage{algpseudocode}
\algnewcommand\algorithmicinput{\textbf{Input:}}
\algnewcommand\INPUT{\item[\algorithmicinput]}
\algnewcommand\algorithmicoutput{\textbf{Output:}}
\algnewcommand\OUTPUT{\item[\algorithmicoutput]}

\usepackage{comment}
\usepackage[normalem]{ulem}

\newcommand\nc\newcommand

\nc{\vzero}{{\boldsymbol{0}}}
\nc{\vone}{{\boldsymbol{1}}}

\nc{\bfC}{{\boldsymbol C}}
\nc{\Read}{{\sf R}}
\nc{\cost}{{\sf cost}}
\nc{\bfa}{{\boldsymbol a}}  \nc{\bfb}{{\boldsymbol b}}  \nc{\bfd}{{\boldsymbol d}}
\nc{\bfe}{{\boldsymbol e}}  \nc{\bff}{{\boldsymbol f}}  \nc{\bfg}{{\boldsymbol g}}
\nc{\bfh}{{\boldsymbol h}}  \nc{\bfi}{{\boldsymbol i}}  \nc{\bfj}{{\boldsymbol j}}
\nc{\bfk}{{\boldsymbol k}}  \nc{\bfl}{{\boldsymbol l}} 
 \nc{\bfm}{{\boldsymbol m}}
\nc{\bfn}{{\boldsymbol n}}  \nc{\bfo}{{\boldsymbol o}}  \nc{\bfp}{{\boldsymbol p}}
\nc{\bfq}{{\boldsymbol q}}  \nc{\bfr}{{\boldsymbol r}}  \nc{\bfs}{{\boldsymbol s}}
\nc{\bft}{{\boldsymbol t}}  \nc{\bfu}{{\boldsymbol u}}  \nc{\bfv}{{\boldsymbol v}}
\nc{\bfw}{{\boldsymbol w}}  \nc{\bfx}{{\boldsymbol x}}  \nc{\bfy}{{\boldsymbol y}}
\nc{\bfz}{{\boldsymbol z}}

\nc\bfA{{\boldsymbol A}}  \nc\bfB{{\boldsymbol B}}  \nc\bfD{{\boldsymbol D}}
\nc\bfE{{\boldsymbol E}}  \nc\bfF{{\boldsymbol F}}  \nc\bfG{{\boldsymbol G}}
\nc\bfH{{\boldsymbol H}}  \nc\bfI{{\boldsymbol I}}  \nc\bfJ{{\boldsymbol J}}
\nc\bfK{{\boldsymbol K}}  \nc\bfL{{\boldsymbol L}}  \nc\bfM{{\boldsymbol M}}
\nc\bfN{{\boldsymbol N}}  \nc\bfO{{\boldsymbol O}}  \nc\bfP{{\boldsymbol P}}
\nc\bfQ{{\boldsymbol Q}}  \nc\bfR{{\boldsymbol R}}  \nc\bfS{{\boldsymbol S}}
\nc\bfT{{\boldsymbol T}}  \nc\bfU{{\boldsymbol U}}  \nc\bfV{{\boldsymbol V}}
\nc\bfW{{\boldsymbol W}}  \nc\bfX{{\boldsymbol X}}  \nc\bfY{{\boldsymbol Y}}
\nc\bfZ{{\boldsymbol Z}}
\nc{\bfc}{{\boldsymbol c}}

\nc\cA{\mathcal{A}}  \nc\cB{\mathcal{B}}  \nc\cC{\mathcal{C}}
\nc\cD{\mathcal{D}}  \nc\cE{\mathcal{E}}  \nc\cF{\mathcal{F}}
\nc\cG{\mathcal{G}}  \nc\cH{\mathcal{H}}  \nc\cI{\mathcal{I}}
\nc\cJ{\mathcal{J}}  \nc\cK{\mathcal{K}}  \nc\cL{\mathcal{L}}
\nc\cM{\mathcal{M}}  \nc\cN{\mathcal{N}}  \nc\cO{\mathcal{O}}
\nc\cP{\mathcal{P}}  \nc\cQ{\mathcal{Q}}  \nc\cR{\mathcal{R}}
\nc\cS{\mathcal{S}}  \nc\cT{\mathcal{T}}  \nc\cU{\mathcal{U}}
\nc\cV{\mathcal{V}}  \nc\cW{\mathcal{W}}  \nc\cX{\mathcal{X}}
\nc\cY{\mathcal{Y}}  \nc\cZ{\mathcal{Z}}

\nc\bbA{\mathbb{A}}  \nc\bbB{\mathbb{B}}  \nc\bbC{\mathbb{C}}
\nc\bbD{\mathbb{D}}  \nc\bbE{\mathbb{E}}  \nc\bbF{\mathbb{F}}
\nc\bbG{\mathbb{G}}  \nc\bbH{\mathbb{H}}  \nc\bbI{\mathbb{I}}
\nc\bbJ{\mathbb{J}}  \nc\bbK{\mathbb{K}}  \nc\bbL{\mathbb{L}}
\nc\bbM{\mathbb{M}}  \nc\bbN{\mathbb{N}}  \nc\bbO{\mathbb{O}}
\nc\bbP{\mathbb{P}}  \nc\bbQ{\mathbb{Q}}  \nc\bbR{\mathbb{R}}
\nc\bbS{\mathbb{S}}  \nc\bbT{\mathbb{T}}  \nc\bbU{\mathbb{U}}
\nc\bbV{\mathbb{V}}  \nc\bbW{\mathbb{W}}  \nc\bbX{\mathbb{X}}
\nc\bbY{\mathbb{Y}}  \nc\bbZ{\mathbb{Z}}

\nc\norm[1]{\left\lVert#1\right\rVert}  % norm: ‖x‖
\nc\setof[1]{\left\{ #1 \right\}}  % set brackets: {x}

\nc\ceil[1]{\left\lceil#1\right\rceil}              % ⌈x⌉
\nc\floor[1]{\left\lfloor#1\right\rfloor}           % ⌊x⌋

\nc\entropy{{\sf H}}        % entropy
\nc\dist{{\rm d}}           % distance
\nc\wt{{\sf wt}}            % Hamming weight

\title{Analytic Combinatorial Method for Coded Coupon Collection}

\author{
  \IEEEauthorblockN{
Jiacheng Cao\IEEEauthorrefmark{1},
Han Mao Kiah\IEEEauthorrefmark{2}}\\[1mm]
\IEEEauthorblockA{\small 
 \IEEEauthorrefmark{1}
School of the Gifted Young, University of Science and Technology of China, China}\\
 \IEEEauthorblockA{\small 
 \IEEEauthorrefmark{2}School of Physical and Mathematical Sciences, Nanyang Technological University, Singapore.
    }
    
  {\footnotesize stanleycao@mail.ustc.edu.cn, hmkiah@ntu.edu.sg. }
  \vspace{-5mm}
  }

\begin{document}

%\begin{comment}
%\section*{Custom New Commands}
%
%\hanmao{Remember to exclude from submission!}
%\begin{itemize}
%  \item \verb|\norm{x}| → $ \norm{x} $ Delimiters will be adjusted in size!
%  \item \verb|\setof{1,2,3}| → $ \setof{1,2,3} $ Delimiters will be adjusted in size!
%  \item \verb|\ceil{x}| → $ \ceil{x} $, \verb|\floor{x}| → $ \floor{x} $ Delimiters will be adjusted in size!
%  \item \verb|\bfv|, \verb|\bfA| → $ \bfv, \bfA $ Use for vectors and matrices! 
%  \item \verb|\cC|, \verb|\cR| → $ \cC, \cR $
%  \item \verb|\bbF|, \verb|\bbZ| → $ \bbF, \bbZ $
%  \item \verb|\entropy| → $ \entropy $, \verb|\wt| → $ \wt $, \verb|\dist| → $ \dist $
%  \item \verb|\hanmao{...}| → color-coded comment (for drafting). Add name and color accordingly!
%  \item \verb|\jiacheng{...}| 
%
%\end{itemize}
%
%\newpage
%\end{comment}

\newpage
\maketitle
\begin{abstract}
We analyse the random-access problem for linear codes under nonuniform sampling using an analytic-combinatorial approach. 
Combining weighted generating functions with the Laplace–Borel transform, we obtain expressions for the expected number of samples required to recover a specified information symbol. 
For systematic MDS codes, we derive an exact formula under arbitrary systematic/parity sampling ratios and study two asymptotic regimes. When the redundancy is fixed, no fixed systematic multiplicity remains optimal as the dimension grows. When the redundancy grows linearly with the dimension, we determine the limiting normalized recovery time and characterize its unique optimal sampling ratio. 
For a singleton–pair construction based on the oriented incidence matrix of the complete graph, we obtain an exact finite-length expression using connected-graph polynomials, including an explicit rational form. We further derive the first finite-size correction to the previously known asymptotic limit. These results provide a means to evaluate recovery performance and optimize sampling allocations at finite blocklengths.
\end{abstract}

\section{Introduction}
\label{sec:intro}

In DNA-based data storage, digital information is encoded into synthetic DNA strands and stored in an unordered pool (see~\cite{SabaryKiahSiegelYaakobi2024} for a survey of coding problems). 
Retrieving a particular piece of information may require repeatedly sampling and sequencing strands, 
making the expected number of samples a natural measure of retrieval cost. 
Bar-Lev et al.~\cite{BarLev2025} introduced a mathematical model for this {\em random-access problem}, 
in which encoded symbols are sampled independently with replacement until the requested information symbol can be recovered. 
The model is closely related to the classical {\em coupon-collector problem}, 
but coding introduces additional possibilities: an information symbol may be recovered from a combination of sampled columns even if it has never been observed directly. 
This raises a natural question: how do the code structure and sampling probabilities affect the expected recovery time?

Two specific questions arise. 
Given a generator matrix and a sampling distribution over its columns, 
we first wish to evaluate the expected recovery time of each information symbol. 
Next, we ask how the sampling probabilities should be chosen to minimize the worst expected recovery time over all information symbols. 
Nonuniform sampling allows different columns, or groups of columns, to be sampled at different rates. 
Understanding how these probabilities affect recovery time is therefore essential for comparing sampling allocations, particularly at finite blocklengths. 
In this paper, we focus on optimizing the sampling distribution for a given generator matrix, 
rather than optimizing the code itself. 
Both problems are formulated precisely in Section~\ref{sec:random-access}.

A natural starting point is to identify the collections of encoded symbols that permit recovery. 
Gruica et al.~\cite{Gruica2025} pursued this combinatorial approach, expressing expected recovery times through the enumeration of recovery sets. 
Their formulas provide exact finite-length evaluations and have been applied to classical codes and constructions from finite geometry~\cite{GruicaMontanucciZullo2026}. 
They also provide a basis for studying repeated columns and nonuniform sampling. 
Boruchovsky et al.~\cite{Boruchovsky2026} and Bodur et al.~\cite{Bodur2026} used related ideas to investigate code constructions and optimize sampling allocations. 
More recently, Wang and Yaakobi~\cite{WangYaakobi2026} developed efficient evaluation algorithms and used distributions over column types to search for improved codes. 
Nevertheless, exact formulas are not always easy to work with. 
Their dependence on the sampling probabilities can be difficult to discern, 
complicating both asymptotic analysis and optimization.

A different perspective has emerged in two recent works on {\em symmetric codes}, where the sampling probability of a column's support depends only on its size.
Hofmeister et al.~\cite{HofmeisterBitarYaakobi2026} observed a connection between fully symmetric binary codes and fountain coding, allowing them to apply techniques from peeling decoding to the asymptotic analysis of random access. 
For codes with singleton and pair columns, they obtained a dilogarithm expression for the limiting normalized recovery time and determined the optimal singleton sampling mass. 
Bar-Lev et al.~\cite{BarLevElishcoGabrys2026} subsequently established the same leading limit under optimal decoding in a broader model over arbitrary fields with independently sampled nonzero coefficients. 
These results provide important asymptotic benchmarks. 
However, the leading limit alone does not reveal how recovery times approach it, nor how the optimal sampling probabilities behave at finite blocklengths.

Our approach draws on the classical analytic treatment of coupon-collection problems developed by Flajolet, Gardy, and Thimonier~\cite{FlajoletGardyThimonier1992}. 
Their framework uses generating functions and the Laplace--Borel transform to obtain integral representations for expected waiting times, including under nonuniform sampling. 
For coded random access, the challenge is to represent the recovery condition in a form that makes these integrals tractable. 
Indeed, while general integral expressions may be difficult to evaluate directly, suitable combinatorial representations can lead to considerable simplifications. 
Building on the recovery-set formulation of Gruica et al.~\cite{Gruica2025}, we use weighted generating functions to describe recovery and non-recovery events. 
Applying the Laplace--Borel transform then yields integral expressions for expected recovery times under nonuniform sampling. 
By exploiting the structure of particular code families, we obtain exact formulas, asymptotic expansions, and sampling-optimization results. We develop this framework in Section~\ref{sec:ac-framework} and illustrate its use on two code families.

We first consider systematic MDS codes in Section~\ref{sec:mds}. 
Gruica et al.~\cite{Gruica2025} had previously studied these codes with repeated systematic columns, obtaining exact recovery-time formulas but leaving the optimal number of repetitions unresolved. 
We allow arbitrary positive sampling ratios between systematic and parity columns and derive an exact expression for the expected recovery time. 
We then examine two asymptotic regimes. 
When the redundancy is fixed and the dimension grows, no fixed systematic multiplicity remains optimal. 
When the redundancy grows linearly with the dimension, we obtain a limiting expression for the normalized recovery time and establish the existence and uniqueness of its optimal sampling ratio. 

In Section~\ref{sec:complete-graph}, we turn to a singleton--pair construction whose parity columns are the oriented incidence vectors of the complete graph. 
Recoverability is governed by the connected components of the sampled graph and the singleton columns observed within them. 
Using connected-graph polynomials, we obtain an exact finite-length expression for the expected recovery time, including an explicit rational form. 
We then derive the first finite-size correction to the asymptotic limit established by Hofmeister et al.~\cite{HofmeisterBitarYaakobi2026} and Bar-Lev et al.~\cite{BarLevElishcoGabrys2026}. 
Even at dimension $k=80$, the distinction between finite-length and asymptotic optimization remains apparent.
For this construction, the numerically optimized normalized recovery time is approximately $0.806854$, 
compared with the asymptotic optimum $0.793933$. 
More notably, the optimal singleton sampling mass at $k=80$ is approximately $0.224070$, 
still appreciably different from its limiting value $0.155474$. 
Thus, even when the recovery time is close to its asymptotic limit, the sampling allocation that minimizes it may remain quite different. 

Our finite-length analysis captures this distinction, allowing both recovery performance and sampling allocation to be studied beyond the leading asymptotic term.

\section{Random-Access Problem}
\label{sec:random-access}

\subsection{Problem Formulation}

Let $\bfG=(\bfg_1,\ldots,\bfg_n)\in
\mathbb{F}_q^{k\times n}$ be a generator matrix of rank $k$.
An information vector $\bfx=(x_1,\ldots,x_k)$ is encoded
as $\bfy=\bfx\bfG$, where $y_j=\bfx\bfg_j$ is the
$j$-th encoded symbol. 
We call $\bfG$ {\em systematic} if its first $k$
columns form the identity matrix $\bfI_k$. In this case,
we write $\bfG=(\bfI_k\mid\bfA)$, where the first $k$
columns are the {\em systematic columns} and the
remaining $n-k$ columns are the {\em parity columns}.

We consider a noiseless sampling process in which
the encoded symbols are drawn independently with replacement.
Let $\bfp=(p_1,\ldots,p_n)$ be the {\em sampling distribution},
where $p_j>0$ for all $j\in[n]$ and
$\sum_{j=1}^n p_j=1$.
For each $t\geq1$, let $J_t$ denote the index of the
symbol obtained in the $t$-th draw, with
$\Pr(J_t=j)=p_j$.
The set of distinct indices observed after $N$ draws is
$R_N=\{J_1,\ldots,J_N\}$, where $R_0=\varnothing$.

The {\em random-access problem} concerns recovering
one requested information symbol $x_i$ or systematic column.
We call $R\subseteq[n]$ a {\em recovery set} for
$x_i$ if
$\bfe_i\in\operatorname{span}_{\mathbb{F}_q}
\{\bfg_j:j\in R\}$,
where $\bfe_i$ is the $i$-th standard basis vector.
Indeed, this condition is equivalent to expressing $x_i$
as a linear combination of the encoded symbols
indexed by $R$.

For $i\in[k]$, define the {\em random-access recovery time}
as
\[
T_i(\bfG,\bfp)
=
\min\left\{
N\geq0:
\bfe_i\in\operatorname{span}_{\mathbb{F}_q}
\{\bfg_j:j\in R_N\}
\right\}.
\]
The quantity $\mathbb{E}[T_i(\bfG,\bfp)]$ is the
{\em expected recovery time} of $x_i$, and
$T_{\max}(\bfG,\bfp)
=\max_{i\in[k]}\mathbb{E}[T_i(\bfG,\bfp)]$
is the {\em maximum expected recovery time}.

We consider the following two problems.

\begin{problem}[Evaluation of Recovery Time]
\label{prob:random-access-evaluation}
Given a generator matrix $\bfG$ and a sampling
distribution $\bfp$, determine
$\mathbb{E}[T_i(\bfG,\bfp)]$ for each $i\in[k]$,
and hence $T_{\max}(\bfG,\bfp)$.
\end{problem}

To formulate the corresponding optimization problem,
let $\mathcal{P}(\bfG)$ be a nonempty family of
{\em admissible sampling distributions} for $\bfG$.
Unless otherwise specified, these distributions are
strictly positive. The family $\mathcal{P}(\bfG)$
may incorporate constraints on the sampling probabilities
or be specified by a collection of parameters.

\begin{problem}[Optimal Sampling Distribution]
\label{prob:random-access-optimization}
Given a generator matrix $\bfG$ and a family
$\mathcal{P}(\bfG)$ of admissible sampling distributions,
determine
\[
T_{\max}^{\star}(\bfG;\mathcal{P})
=
\inf_{\bfp\in\mathcal{P}(\bfG)}
T_{\max}(\bfG,\bfp),
\]
and characterize an optimal sampling distribution,
when one exists.
\end{problem}

The two problems distinguish evaluation from optimization.
In Problem~\ref{prob:random-access-evaluation},
both the generator matrix and the sampling distribution
are fixed. In Problem~\ref{prob:random-access-optimization},
we retain the generator matrix but allow the sampling
distribution to vary within an admissible family.

In this paper, we focus on systematic generator matrices
and sampling distributions that assign the same probability
to columns of the same type. Specifically, each systematic
column is sampled with probability $p_{\rm sys}$, while
each parity column is sampled with probability $p_{\rm par}$,
where
$kp_{\rm sys}+(n-k)p_{\rm par}=1$.
We define the {\em sampling ratio} as
\[
\sigma\triangleq \frac{p_{\rm par}}{p_{\rm sys}}.
\]
Thus, the sampling distribution is uniquely determined
by $\sigma>0$. We investigate the expected recovery time
and the optimal sampling ratio for two families of codes.

%In this paper, we focus on systematic generator matrices
%and sampling distributions that assign the same
%probability to columns of the same type. Specifically,
%each systematic column is sampled with probability
%$p_{\mathrm{sys}}$, while each parity column is sampled
%with probability $p_{\mathrm{par}}$, where
%$kp_{\mathrm{sys}}+(n-k)p_{\mathrm{par}}=1$.
%
%Thus the sampling distribution is determined by a
%single parameter. We will study how this parameter
%affects the expected recovery time and seek its
%optimal value for two families of codes.

\subsection{Realizing Sampling Distributions}
\label{subsec:sampling-realization}

In DNA storage, nonuniform sampling distributions
can be realized by varying the relative concentrations
of encoded strands. More broadly, controlling
probabilities through molecular concentrations is an
established idea in DNA computing. Adar et al.
\cite{Adar2004} experimentally demonstrated a
stochastic biomolecular automaton whose transition
probabilities were controlled by the relative
concentrations of competing DNA molecules.
Related ideas arise in concentration programming
for algorithmic self-assembly; see the survey
by Doty \cite{Doty2012}.

To relate concentrations to our sampling model,
suppose that the $j$-th encoded strand has concentration
$c_j>0$. Under uniform sampling from the DNA pool,
its sampling probability is
$p_j=c_j/\sum_{\ell=1}^n c_\ell$.
For a systematic generator matrix, suppose that all
systematic strands have concentration $c_{\rm sys}$
and all parity strands have concentration $c_{\rm par}$.
Then $p_{\rm par}/p_{\rm sys}
=c_{\rm par}/c_{\rm sys}$.
Thus the relative sampling probabilities can be
controlled by adjusting the relative concentrations.

Alternatively, rational sampling distributions can be
realized by repeating columns of the generator matrix
and sampling uniformly from the enlarged matrix,
as considered in \cite{Gruica2025,Boruchovsky2026}.

\subsection{Previous Results}
\label{subsec:previous-results}

Bar-Lev et al. \cite{BarLev2025} introduced the
DNA coverage depth problem, which studies the number
of sequencing reads needed to recover stored information.
They considered both the recovery of the entire
information vector and what they called the
{\em singleton-random-access problem}, which is
later simply referred to as the {\em random-access problem}.
For the latter, they showed that systematic MDS codes
and simple parity codes have expected recovery time $k$
under uniform sampling. They also established general
lower bounds on the maximum expected recovery time.

Several subsequent works addressed the evaluation
problem (Problem~\ref{prob:random-access-evaluation}).
Gruica et al. \cite{Gruica2025} derived general
combinatorial formulas for computing recovery times
under uniform sampling, based on counting subsets
of columns that permit recovery. They introduced
{\em recovery-balanced codes} and proved that
systematic generator matrices of such codes have
$T_{\max}=k$. These include MDS codes and codes
with transitive permutation automorphism groups,
such as binary simplex and Hamming codes.

In contrast, finite-geometric constructions can
achieve expected recovery times below $k$ even under
uniform sampling. Gruica, Montanucci, and Zullo
\cite{GruicaMontanucciZullo2026} introduced
{\em balanced quasi-arcs} and derived recovery-time
formulas for the corresponding generator matrices.
Bodur et al. \cite{Bodur2026} obtained exact
asymptotic expressions for certain constructions
and studied the full distribution of the recovery
time. More recently, Van de Voorde and Zullo
\cite{VanDeVoordeZullo2026} investigated weak arcs,
obtaining explicit recovery-time formulas for
several geometric constructions in projective
planes and three-dimensional projective spaces.
For general generator matrices, Wang and Yaakobi
\cite{WangYaakobi2026} developed an algorithm
that computes the exact expectation in $O(n)$ time
when $q$ and $k$ are fixed.

Nonuniform sampling has also been studied through
the use of repeated columns. For example,
Gruica et al. \cite{Gruica2025} considered
{\em perturbed} MDS and simplex codes, obtained
by repeating each systematic column $x$ times
while retaining one copy of each parity column.
In our formulation, this corresponds to
$p_{\rm sys}/p_{\rm par}=x$, where $x$ is a
positive integer. They derived exact finite-length
formulas for the expected recovery times, although
the optimal value of $x$ was not determined.
Similar techniques were considered for balanced
quasi-arcs \cite{GruicaMontanucciZullo2026}
and weak arcs \cite{VanDeVoordeZullo2026},
where additional copies of the fundamental points
were used to reduce the expected recovery time.

Several works have also investigated the optimization problem (Problem~\ref{prob:random-access-optimization}).
Boruchovsky et al.~\cite{Boruchovsky2026} formulated an optimization over probability distributions on projective points and solved the case $k=2$.
For larger $k$, they constructed recovery-complete generator matrices using weight-two columns whose nonzero coefficients were selected using $B_{k-1}$ sequences.
They also studied the effect of repeating the systematic columns.
For balanced quasi-arcs, Bodur et al.~\cite{Bodur2026} derived an asymptotic expression for the recovery time and determined the optimal ratio of systematic multiplicity to the number of non-systematic points on each side of the fundamental triangle.
For binary fully symmetric codes with singleton and pair columns, Hofmeister et al.~\cite[Section~V-D]{HofmeisterBitarYaakobi2026} derived the asymptotic recovery time under peeling decoding, expressed it in terms of the dilogarithm, and determined the optimal singleton sampling mass.
Bar-Lev et al.~\cite[Proposition~12]{BarLevElishcoGabrys2026} subsequently established the same leading asymptotic limit under optimal decoding in a more general singleton--pair model over arbitrary fields with nonzero coefficient distributions.
Finally, Wang and Yaakobi~\cite{WangYaakobi2026} reformulated the search for generator matrices in terms of distributions over column types, enabling numerical optimization for small parameters.

Table~\ref{tab:previous-results} summarizes
these results in terms of the generator matrices
and sampling distributions considered. We
distinguish uniform sampling from distributions
obtained by repeating columns, even when both
settings are studied in the same paper.

\begin{table*}[!t]
\centering
\caption{Previous results on the random-access problem,
classified by generator matrix and sampling distribution.}
\label{tab:previous-results}
\small
\setlength{\tabcolsep}{2.5pt}
\renewcommand{\arraystretch}{1.12}
\begin{tabular}{
  p{0.20\textwidth}
  p{0.26\textwidth}
  p{0.20\textwidth}
  p{0.26\textwidth}
}
\hline
Work &
Generator matrix $\bfG$ &
Sampling family $\mathcal P$ &
Main results \\
\hline

Bar-Lev et al.
\cite{BarLev2025}
&
Identity, simple parity, systematic MDS,
and other constructions
&
Uniform
&
Baseline recovery-time evaluations and
general lower bounds.
\\
\hline

Gruica et al.
\cite{Gruica2025}
&
Systematic MDS, simplex, and other
recovery-balanced codes
&
Uniform
&
$T_{\max}=k$ for systematic
recovery-balanced codes; general
combinatorial formulas.
\\
\cline{2-4}
&
Perturbed systematic MDS and simplex
&
$p_{\rm sys}/p_{\rm par}=x$,\newline
integer $x\geq 1$
&
Exact finite-length formulas
(Theorems 2--3); optimal $x$
not determined.
\\
\hline

Gruica--Montanucci--Zullo
\cite{GruicaMontanucciZullo2026}
&
Balanced quasi-arcs in
$\mathrm{PG}(2,q)$
&
Uniform
&
Recovery-time formulas and
constructions with $T_{\max}<3$.
\\
\cline{2-4}
&
Balanced quasi-arcs with repeated
fundamental points
&
$p_{\rm sys}/p_{\rm par}=y$,
integer $y\geq 1$
&
Exact formulas; asymptotic bound
for $y=x$, where $x$ is the
number of points per side.
\\
\hline

Bodur et al.
\cite{Bodur2026}
&
Rate-half constructions
&
Uniform
&
Exact asymptotic recovery time.
\\
\cline{2-4}
&
General linear codes&
Uniform
&
Higher moments and distribution.
\\
\cline{2-4}
&
Balanced quasi-arcs with $x$ points
per side and $y$ copies of each
fundamental point
&
$p_{\rm sys}/p_{\rm par}=y$
&
Exact asymptotic expression and
optimal ratio $y/x$.
\\
\hline

Boruchovsky et al.
\cite{Boruchovsky2026}
&
Projective points in
$\mathrm{PG}(1,q)$, $k=2$
&
Arbitrary distributions
&
Exact optimal distribution and
recovery time.
\\
\cline{2-4}
&
$\bfI_k$ with $x$ distinct
weight-two columns per pair,
of the form
$\bfe_i+\beta^a\bfe_j$
&
Uniform
&
Recovery-complete construction
using $B_{k-1}$ sequences.
\\
\cline{2-4}
&
Same construction with $y$
copies of each systematic column
&
\(p_{\mathrm{sys}}/p_{\mathrm{par}}=y\), asymptotic parameter \(y/x\)
&
Asymptotic recovery-time analysis
and upper bounds.
\\
\hline

Wang--Yaakobi
\cite{WangYaakobi2026}
&
Arbitrary generator matrices
&
Uniform
&
Exact evaluation algorithm in
$O(n)$ time for fixed $q,k$.
\\
\cline{2-4}
&
Generator matrices with varying
column multiplicities
&
Distributions over column types,
including Hamming-weight classes
&
Optimization over weight
distributions and improved
constructions for small $k$.
\\
\hline

Van de Voorde--Zullo
\cite{VanDeVoordeZullo2026}
&
Planar weak arcs and tetrahedral
configurations
&
Uniform
&
Explicit recovery-time formulas
and optimality within restricted
geometric families.
\\
\cline{2-4}
&
Tetrahedral constructions with
repeated fundamental points;
other weighted projective sets
&
Multiplicity-induced\newline
distributions
&
Exact formulas and weighted
constructions (Propositions 8--10).
\\
\hline

Hofmeister et al.~\cite{HofmeisterBitarYaakobi2026}
& Binary fully symmetric singleton--pair codes
& Singleton–pair allocation\newline \(\alpha=kp_{\mathrm{sys}}\)
& Peeling-decoding asymptotic; dilogarithm expression
  and optimal singleton mass. \\
\hline
Bar-Lev et al.~\cite{BarLevElishcoGabrys2026}
& Singleton--pair ensembles over arbitrary fields
  with nonzero coefficients 
& Total singleton-support mass\newline \(\alpha\)
& Same leading asymptotic limit under optimal
  decoding (Proposition~12). \\
\hline
\end{tabular}
\end{table*}

Related extensions include the simultaneous recovery
of multiple information symbols
\cite{GruicaPetrilloZullo2026} and block-structured
information retrieval \cite{BarLevCodedRetrieval}.
These consider recovery objectives different from
the single-symbol problem studied here.

\subsection{Our Contributions}

In this paper, we develop an analytic combinatorial approach to evaluating and optimizing the expected recovery time under nonuniform sampling. Our main contributions are summarized as follows.

\begin{enumerate}[(A)]
    \item \textbf{Analytic combinatorial framework.}
    We extend the combinatorial approach of Gruica et al.~\cite{Gruica2025} to nonuniform sampling using weighted exponential generating functions.
    Applying the Laplace--Borel transform yields an integral representation for the expected recovery time in terms of recovery or non-recovery sets.
    This framework provides a common starting point for the analysis of different code constructions.

    \item \textbf{Systematic MDS codes.}
    We derive an exact expression for the expected recovery time of systematic MDS codes under arbitrary systematic/parity sampling ratios.
    For fixed redundancy, we obtain an asymptotic expansion showing that no fixed multiplicity of the systematic columns remains optimal as the dimension grows.
    When the redundancy grows linearly with the dimension, we determine the limiting normalized recovery time and establish the existence and uniqueness of the optimal sampling ratio.

    \item \textbf{Singleton--pair incidence construction.}
    We study the construction whose parity columns are the oriented incidence vectors of the complete graph.
    Using connected-graph polynomials, we derive an exact finite-length expression for the expected recovery time and show that it is a rational function of the sampling probabilities.
    We further obtain a second-order asymptotic expansion, identifying the first finite-size correction to the leading limit established by Hofmeister et al.~\cite{HofmeisterBitarYaakobi2026} and Bar-Lev et al.~\cite{BarLevElishcoGabrys2026}.
    This correction provides an approximation to the optimal sampling allocation at finite blocklengths.
\end{enumerate}

\section{Analytic Combinatorial Framework}
\label{sec:ac-framework}

We begin with a general framework for coded coupon collection.
Gruica et al.~\cite{Gruica2025} established a combinatorial
approach that expresses the expected recovery time in terms
of the numbers of recovery sets of each cardinality.
Building on this approach, we use weighted exponential
generating functions to incorporate non-uniform sampling
distributions. The Laplace--Borel transform then converts
the resulting discrete sums into integral representations,
which are amenable to exact evaluation and asymptotic
analysis. This follows the classical analytic-combinatorial
methods of Flajolet et al.~\cite{FlajoletGardyThimonier1992};
see also \cite[Sec.~II.3.2]{FlajoletSedgewick2009}.

%We begin with a general framework for coded coupon collection. The
%purpose of this section is to separate the combinatorial description of
%recovery from the sampling process. Once the family of recovery sets is
%specified, the distribution and expectation of the corresponding
%collection time can be encoded using exponential generating functions.
%The expected collection time is then obtained by a Laplace--Borel
%transform.
%
%This point of view will be useful in two rather different settings. For
%MDS codes, non-recovery can be described by a simple threshold condition
%on the numbers of systematic and parity coupons that have been
%collected. For the complete-graph incidence codes studied later, the
%non-recovery sets are naturally described in terms of connected
%components of a graph. In both cases, the same analytic framework
%applies.

\subsection{Recovery and non-recovery generating functions}

Consider $n$ coupon types indexed by
$[n]=\{1,\ldots,n\}$. Coupon $i$ is drawn
independently with probability $p_i>0$, where
$\sum_{i=1}^n p_i=1$. We fix an object that
we wish to recover from the observed coupons.

For a subset $R\subseteq[n]$, define the
recovery indicator $\mathbb I(R)$ to be one
if the coupons indexed by $R$ suffice for
recovery, and zero otherwise. We assume that
$\mathbb I(\varnothing)=0$,
$\mathbb I([n])=1$, and that collecting
additional coupons cannot invalidate recovery.
In the coding-theoretic setting of
Section~\ref{sec:random-access}, where the
requested information symbol is $x_i$,
the recovery condition is
$\bfe_i\in\operatorname{span}_{\mathbb F_q}
\{\bfg_j:j\in R\}$.

Let $R_N\subseteq[n]$ denote the set of
distinct coupon types observed among the
first $N$ draws, with $R_0=\varnothing$.
The collection time $T$ is the smallest
$N$ for which $\mathbb I(R_N)=1$.
We first encode the recovery sets.

%Consider $n$ coupon types indexed by $[n]=\{1,\ldots,n\}$. Coupon $i$
%is drawn independently with probability $p_i>0$, where
%$\sum_{i=1}^n p_i=1$. We fix some object that we wish to recover from
%the coupons that have been observed.
%
%For a subset $R\subseteq[n]$, let $\mathbb{I}(R)$ denote the recovery
%indicator, where $\mathbb{I}(R)=1$ if the coupons indexed by $R$ suffice
%for recovery, and $\mathbb{I}(R)=0$ otherwise. In the coding-theoretic
%setting, for example, one may have a generator matrix with columns
%$g_1,\ldots,g_n$, and $\mathbb{I}(R)=1$ precisely when the desired
%information symbol belongs to the span of $\{g_i:i\in R\}$.
%
%Let $S_N\subseteq[n]$ denote the set of distinct coupon types observed
%among the first $N$ draws, with $S_0=\varnothing$. We define the
%collection time $T$ to be the smallest $N$ for which
%$\mathbb{I}(S_N)=1$. We assume throughout that recovery is eventually
%possible, or equivalently that $\mathbb{I}([n])=1$.

\begin{definition}[Recovery EGF]
\label{def:recovery-egf}
The recovery exponential generating function is
$F(z)=\sum_{R\subseteq[n]}\mathbb{I}(R)
\prod_{i\in R}(e^{p_i z}-1)$.
\end{definition}

The factors $e^{p_i z}-1$ have a simple coupon-collection
interpretation. For a fixed set $R$, the product
$\prod_{i\in R}(e^{p_i z}-1)$ encodes sequences of draws whose set of
distinct coupon types is exactly $R$.

More precisely, expanding the product gives
$\prod_{i\in R}(e^{p_i z}-1)
=\sum_{(m_i:i\in R),\,m_i\geq 1}
\prod_{i\in R}(p_i z)^{m_i}/m_i!$.
Hence the coefficient of $z^N/N!$ is the sum, over all positive
integers $(m_i:i\in R)$ satisfying $\sum_{i\in R}m_i=N$, of
$N!\prod_{i\in R}p_i^{m_i}/\prod_{i\in R}m_i!$. This is precisely the
probability that, among the first $N$ draws, coupon $i$ appears exactly
$m_i$ times for every $i\in R$, while no coupon outside $R$ appears.
Summing over all such multiplicities therefore gives the probability
that $R_N=R$.

This immediately yields the following coefficient formula.

\begin{proposition}
\label{prop:recovery-egf}
For every $N\geq 0$, we have
$\Pr(T\leq N)=N![z^N]F(z)$.
\end{proposition}

\begin{proof}
For every $R\subseteq[n]$, the preceding discussion shows that
$N![z^N]\prod_{i\in R}(e^{p_i z}-1)=\Pr(R_N=R)$. Therefore
$N![z^N]F(z)$ is the sum of $\Pr(R_N=R)$ over all sets $R$ satisfying
$\mathbb{I}(R)=1$. This is exactly the probability that the coupons
observed by time $N$ permit recovery, namely $\Pr(T\leq N)$.
\end{proof}

For the applications in this paper, it will often be more convenient
to enumerate the sets that do \emph{not} permit recovery.

\begin{definition}[Non-recovery EGF]
\label{def:nonrecovery-egf}
The non-recovery exponential generating function is
$\overline{F}(z)=\sum_{R\subseteq[n]}(1-\mathbb{I}(R))
\prod_{i\in R}(e^{p_i z}-1)$.
\end{definition}

The recovery and non-recovery EGFs are complementary in a particularly
simple way. Indeed,
$F(z)+\overline{F}(z)
=\sum_{R\subseteq[n]}\prod_{i\in R}(e^{p_i z}-1)$.
Summing independently over whether each coupon type belongs to $R$
gives
$F(z)+\overline{F}(z)
=\prod_{i=1}^n(1+(e^{p_i z}-1))
=\prod_{i=1}^n e^{p_i z}
=e^z$.

We therefore obtain the following counterpart of
Proposition~\ref{prop:recovery-egf}.

\begin{corollary}
\label{cor:nonrecovery-egf}
The recovery and non-recovery EGFs satisfy
$\overline{F}(z)=e^z-F(z)$. Moreover, for every $N\geq 0$,
$\Pr(T>N)=N![z^N]\overline{F}(z)$.
\end{corollary}

\begin{proof}
The identity $\overline{F}(z)=e^z-F(z)$ follows from the calculation
above. The coefficient identity can either be proved directly, by
summing over the sets $R$ for which $\mathbb{I}(R)=0$, or by using
Proposition~\ref{prop:recovery-egf} together with
$\Pr(T>N)=1-\Pr(T\leq N)$.
\end{proof}

%Thus $F$ naturally encodes the cumulative distribution of $T$, whereas
%$\overline{F}$ naturally encodes its survival probabilities. This
%distinction will be useful later: in many coded coupon-collection
%problems, describing the non-recovery sets is substantially easier than
%describing all recovery sets.

\subsection{The Laplace--Borel representation}

By the tail-sum formula,
$\mathbb E[T]=\sum_{N\geq 0}\Pr(T>N)$ and 
the coefficients of the non-recovery EGF are precisely
these probabilities. The Laplace--Borel transform then allows
us to recover their sum directly from the EGF.

The Laplace--Borel transform provides precisely this
conversion. Indeed, for every integer $N\geq 0$,
\[
\int_0^\infty e^{-t}\frac{t^N}{N!}\,dt=1,
\]
which follows from the classical gamma integral
$\Gamma(N+1)=N!$. Consequently, if
$G(z)=\sum_{N\geq 0}g_Nz^N/N!$ has nonnegative
coefficients, Tonelli's theorem gives
\[
\int_0^\infty e^{-t}G(t)\,dt
=\sum_{N\geq 0}g_N.
\]
This technique is classical in analytic combinatorics;
see, for example,
\cite[Sec.~II.3.2]{FlajoletSedgewick2009}.

Applying this transform to the non-recovery EGF,
and using its complementary relationship with the
recovery EGF, yields two equivalent integral
representations for the expected collection time.

\begin{theorem}[Laplace--Borel representation]
\label{thm:laplace-borel}
Let $T$ be the coded coupon-collection time defined above. Then
$\mathbb{E}[T]=\int_0^\infty e^{-t}\overline{F}(t)\,dt$. Equivalently,
using $\overline{F}(t)=e^t-F(t)$, we have
$\mathbb{E}[T]=\int_0^\infty(1-e^{-t}F(t))\,dt$.
\end{theorem}

\begin{proof}
Since $T$ is a nonnegative integer-valued random variable, the
tail-sum formula gives
$\mathbb{E}[T]=\sum_{N\geq 0}\Pr(T>N)$.
By Corollary~\ref{cor:nonrecovery-egf}, the non-recovery EGF can be
written as
$\overline{F}(t)=\sum_{N\geq 0}\Pr(T>N)t^N/N!$.
All coefficients are nonnegative, so Tonelli's theorem allows us to
interchange the sum and the integral. Hence
$\int_0^\infty e^{-t}\overline{F}(t)\,dt
=\sum_{N\geq 0}\Pr(T>N)
\int_0^\infty e^{-t}t^N/N!\,dt
=\sum_{N\geq 0}\Pr(T>N)
=\mathbb{E}[T]$.
Finally, substituting $\overline{F}(t)=e^t-F(t)$ gives the equivalent
formula
$\mathbb{E}[T]=\int_0^\infty(1-e^{-t}F(t))\,dt$.
\end{proof}

The non-recovery representation is useful in both of our
applications. For MDS codes, non-recovery is characterized
by the numbers of distinct systematic and parity coupons
collected. For complete-graph incidence codes, it is governed
by the connected components induced by the collected parity
coupons.

%The two forms in Theorem~\ref{thm:laplace-borel} are mathematically
%equivalent, but they need not be equally convenient. If recovery sets
%have a simple description, it may be natural to work directly with
%$F$. On the other hand, if failure of recovery has a simpler
%combinatorial structure, then the representation
%$\mathbb{E}[T]=\int_0^\infty e^{-t}\overline{F}(t)\,dt$ is usually more
%useful.
%
%The latter situation occurs in both of our main applications. 
%For MDS codes, non-recovery can be characterized by a threshold on
%the numbers of distinct systematic and parity coupons collected. 
%For complete-graph incidence codes, non-recovery is characterized by the
%connected component containing the requested systematic node having no
%collected systematic coupon. 
%The same Laplace--Borel identity can therefore be applied in both settings, 
%although the underlying combinatorial descriptions are quite different.

\subsection{Uniform sampling}

We conclude this section by specializing the framework to uniform
sampling, thereby recovering the known formula for random access from
Gruica~et~al.~\cite[Lemma~1]{Gruica2025}.

Suppose that $p_i=1/n$ for every $i\in[n]$. For $0\leq s\leq n$, let
$\alpha(s)$ denote the number of recovery sets of cardinality $s$,
that is,
$\alpha(s)=|\{R\subseteq[n]:|R|=s,\ \mathbb{I}(R)=1\}|$.
Since all subsets of the same cardinality now contribute the same
weight, the non-recovery EGF simplifies to
$\overline{F}(z)=\sum_{s=0}^n
(\binom{n}{s}-\alpha(s))(e^{z/n}-1)^s$.

Because the full set $[n]$ permits recovery, we have $\alpha(n)=1$,
and hence the term corresponding to $s=n$ vanishes. Applying
Theorem~\ref{thm:laplace-borel} therefore gives
$\mathbb{E}[T]=\sum_{s=0}^{n-1}
(\binom{n}{s}-\alpha(s))I_{n,s}$, where
$I_{n,s}=\int_0^\infty e^{-t}(e^{t/n}-1)^s\,dt$.

The integral $I_{n,s}$ has a simple beta-integral evaluation. Setting
$x=e^{-t/n}$ gives
$I_{n,s}=n\int_0^1x^{n-s-1}(1-x)^s\,dx
=nB(n-s,s+1)$.
For $0\leq s\leq n-1$, this becomes
$I_{n,s}=n(n-s-1)!s!/n!
=(n-s-1)!s!/(n-1)!
=1/\binom{n-1}{s}$.

Consequently,
$\mathbb{E}[T]=\sum_{s=0}^{n-1}
(\binom{n}{s}-\alpha(s))/\binom{n-1}{s}$.
Using
$\binom{n}{s}/\binom{n-1}{s}=n/(n-s)$, we obtain
$\sum_{s=0}^{n-1}\binom{n}{s}/\binom{n-1}{s}
=n\sum_{j=1}^n1/j=nH_n$.
Since $\alpha(0)=0$, this yields the following corollary.

\begin{corollary}[Uniform sampling]
\label{cor:uniform-sampling}
Under uniform sampling, where $p_i=1/n$ for all $i\in[n]$, the expected
collection time is
$\mathbb{E}[T]=nH_n-\sum_{s=1}^{n-1}
\alpha(s)/\binom{n-1}{s}$.
\end{corollary}

%In the random-access setting of Gruica~et~al., the coupons are the
%columns $g_1,\ldots,g_n$ of a generator matrix and, for a requested
%information symbol $i$, their quantity $\alpha_i(s)$ counts the
%$s$-subsets $R\subseteq[n]$ for which the desired basis vector $\bfe_i$
%belongs to the span of $\{\bfg_j:j\in R\}$. Thus
%Corollary~\ref{cor:uniform-sampling} becomes exactly
%\cite[Lemma~1]{Gruica2025}.
%
%Their derivation proceeds by conditioning on the number of distinct
%columns observed after a given number of draws and then applying
%inclusion--exclusion. The analytic-combinatorial framework above gives
%the same expression directly from the non-recovery EGF, with the
%remaining calculation reduced to a beta integral. More importantly,
%the generating-function formulation does not require uniform sampling,
%which will allow us to study deliberately nonuniform sampling
%distributions in the following sections.
\section{Recovery Time for MDS Codes}
\label{sec:mds}

Let $\bfG=(\bfI_k\mid\bfA)$ be a systematic generator
matrix of an $[n,k]$ MDS code, where $n=k+r$ and
$r\geq1$. For sampling ratio $\sigma>0$, we have
\[
p_{\rm sys}=\frac{1}{k+r\sigma},
\qquad
p_{\rm par}=\frac{\sigma}{k+r\sigma}.
\]

\subsection{Exact Recovery Time}
\label{subsec:mds-exact}

%Fix a systematic symbol $i\in[k]$.
Since $\bfG$ is MDS, the $i$-th systematic symbol can be recovered
either by sampling its corresponding systematic
column or by sampling at least $k$ distinct
columns. In the latter case, suppose that $a$
of the other $k-1$ systematic columns and $b$
of the $r$ parity columns have not been sampled.
Recovery is then possible precisely when
$a+b\leq r-1$.

For convenience, define the index set
$\mathcal D=\{(a,b):0\leq a\leq k-1,\,
0\leq b\leq r,\,a+b\leq r-1\}$.

\begin{theorem}[MDS Recovery Time]
\label{thm:mds-exact}
Let $\bfG$ be a systematic $[k+r,k]$ MDS
generator matrix. For sampling ratio $\sigma>0$,
we have
\[
\mathbb E[T_i]
=(k+r\sigma)\left(
1-\sum_{(a,b)\in\mathcal D}
\binom{k-1}{a}\binom rb
\int_0^1
(1-u)^{k-1-a}(1-u^\sigma)^{r-b}
u^{a+\sigma b}\,du
\right).
\]
The expression is independent of $i$, and
hence gives $T_{\max}(\bfG,\boldsymbol p)$.
\end{theorem}

\begin{proof}
The $i$-th systematic symbol cannot be recovered
precisely when its corresponding systematic column
has not been sampled and fewer than $k$ distinct
other columns have been sampled.

Suppose that $a$ of the other $k-1$ systematic
columns and $b$ of the $r$ parity columns are
missing. Non-recovery then occurs precisely when
$a+b\ge r$. Consequently, the non-recovery EGF is
\[
\overline F_i(z)
=\sum_{\substack{0\le a\le k-1,\;0\le b\le r\\a+b\ge r}}
\binom{k-1}{a}\binom rb
(e^{p_{\rm sys}z}-1)^{k-1-a}
(e^{p_{\rm par}z}-1)^{r-b}.
\]

By Theorem~\ref{thm:laplace-borel},
$\mathbb E[T_i]=\int_0^\infty e^{-t}
\overline F_i(t)\,dt$.
Substituting $u=e^{-p_{\rm sys}t}$ and using
$p_{\rm par}=\sigma p_{\rm sys}$, we obtain
\[
\mathbb E[T_i]
=\frac1{p_{\rm sys}}
\sum_{\substack{0\le a\le k-1,\;0\le b\le r\\a+b\ge r}}
\binom{k-1}{a}\binom rb
\int_0^1
(1-u)^{k-1-a}(1-u^\sigma)^{r-b}
u^{a+\sigma b}\,du.
\]

By the binomial theorem, the sum of the
integrands over all pairs
$0\le a\le k-1$ and $0\le b\le r$,
including their binomial coefficients,
equals
$((1-u)+u)^{k-1}
((1-u^\sigma)+u^\sigma)^r=1$.

Since $\mathcal D$ consists precisely of
the complementary pairs satisfying
$a+b\le r-1$, the preceding expression
reduces to the stated formula upon using
$1/p_{\rm sys}=k+r\sigma$.
\end{proof}

\begin{remark}[Uniform Sampling and Systematic-Only Sampling]
\label{rem:mds-uniform}
When $\sigma=1$, all $n=k+r$ columns are
sampled uniformly. Grouping the summands
in Theorem~\ref{thm:mds-exact} according to
$j=a+b$ and applying Vandermonde's identity
gives
$\sum_{a+b=j}\binom{k-1}{a}\binom rb
=\binom{n-1}{j}$.
Moreover, the corresponding integral equals
$B(j+1,n-j)=1/(n\binom{n-1}{j})$.
Hence,
\[
\mathbb E[T_i]
=n\left(1-\sum_{j=0}^{r-1}\frac1n\right)
=k.
\]

At the other extreme, $\sigma\downarrow0$
corresponds to sampling only systematic
columns. Since $r-b\geq1$ for every
$(a,b)\in\mathcal D$, the factor
$(1-u^\sigma)^{r-b}$ tends to zero.
Thus every integral in the summation
vanishes in the limit, giving
$\lim_{\sigma\downarrow0}\mathbb E[T_i]=k$.
\end{remark}

\begin{remark}[Computational Complexity]
\label{rem:mds-complexity}
Theorem~\ref{thm:mds-exact} involves at most
$r(r+1)/2$ one-dimensional integrals.
If each integral is evaluated using a
quadrature rule with $M$ evaluation points,
the direct evaluation requires
$O(Mr^2)$ arithmetic operations.

For comparison, Gruica et al.
\cite{Gruica2025} obtained an exact formula
for perturbed MDS codes by repeating each
systematic column $x$ times while retaining
one copy of each parity column. Under uniform
sampling of the resulting matrix, the effective
sampling probabilities satisfy
$p_{\rm sys}/p_{\rm par}=x$, or equivalently
$\sigma=1/x$.

Their formula involves four nested summations,
whereas our expression involves at most
$O(r^2)$ one-dimensional integrals.
In particular, the number of integrals does
not depend on the dimension $k$.
\end{remark}

\subsection{Asymptotic Recovery Time for Fixed Redundancy}
\label{subsec:mds-fixed-r}

For fixed redundancy $r$, we obtain the following
asymptotic expansion for the expected recovery time
as $k\to\infty$.

\begin{theorem}[Fixed-Redundancy Asymptotics]
\label{thm:mds-fixed-r}
Fix an integer $r\geq1$ and $0<\sigma<1$.

For a systematic $[k+r,k]$ MDS code
with sampling ratio $\sigma$,

\[
\mathbb E[T_i]
=k-r(1-\sigma)
+\frac{r\Gamma(r+\sigma)}{\Gamma(r)}k^{-\sigma}
+O\bigl(k^{-\min\{2\sigma,1\}}\bigr).
\]
In particular,
$\mathbb E[T_i]=k-r(1-\sigma)+O(k^{-\sigma})$.
\end{theorem}

\begin{proof}
For $u\in[0,1]$, let $A_u$ and $B_u$ be
independent random variables with
$A_u\sim\operatorname{Bin}(k-1,u)$ and
$B_u\sim\operatorname{Bin}(r,u^\sigma)$.
By Theorem~\ref{thm:mds-exact},
\[
\mathbb E[T_i]
=(k+r\sigma)\int_0^1\Pr(A_u+B_u\geq r)\,du.
\]

For sufficiently large $k$, the beta-integral
identity gives
$\int_0^1\Pr(A_u<r)\,du=r/k$.
Define
$\Delta_k=\int_0^1
\bigl(\Pr(A_u<r)-\Pr(A_u+B_u<r)\bigr)\,du$.
It follows that
$\mathbb E[T_i]=(k+r\sigma)(1-r/k+\Delta_k)$.

Conditioning on $B_u$, we obtain
\[
\Delta_k
=\sum_{b=1}^{r}\binom rb
\int_0^1 u^{b\sigma}(1-u^\sigma)^{r-b}
\Pr(r-b\leq A_u\leq r-1)\,du.
\]
We show that the leading contribution comes
from $b=1$.

For any fixed integer $a\geq0$ and $\alpha>0$,
\[
\int_0^1u^\alpha\Pr(A_u=a)\,du
=\frac{\Gamma(k)\Gamma(a+\alpha+1)}
{\Gamma(a+1)\Gamma(k+\alpha+1)}
=O(k^{-1-\alpha}).
\]
Since $r$ is fixed, the terms with $b\geq2$
contribute $O(k^{-1-2\sigma})$ in total.

For $b=1$, the event
$r-b\leq A_u\leq r-1$ reduces to
$A_u=r-1$. Moreover,
$(1-u^\sigma)^{r-1}=1+O(u^\sigma)$.
Therefore,
\[
\Delta_k
=r\int_0^1u^\sigma\Pr(A_u=r-1)\,du
+O(k^{-1-2\sigma})
=\frac{r\Gamma(r+\sigma)\Gamma(k)}
{\Gamma(r)\Gamma(k+1+\sigma)}
+O(k^{-1-2\sigma}).
\]

Using the asymptotic identity
$\Gamma(k)/\Gamma(k+1+\sigma)
=k^{-1-\sigma}(1+O(k^{-1}))$,
we obtain
\[
\Delta_k
=\frac{r\Gamma(r+\sigma)}{\Gamma(r)}
k^{-1-\sigma}+O(k^{-1-2\sigma}).
\]
Finally, substituting this into
$\mathbb E[T_i]=(k+r\sigma)(1-r/k+\Delta_k)$
and collecting terms gives the result.
\end{proof}

\begin{remark}[Multiplicity of Systematic Columns]
\label{rem:mds-simple-parity}
In Remark~5 of \cite{Gruica2025},
Gruica et al. observed experimentally that,
for simple parity codes, the optimal number
$x$ of copies of each systematic column
appeared to be $3$. They noted that this
observation was difficult to prove formally.

However, Theorem~\ref{thm:mds-fixed-r}
shows that $x=3$ is suboptimal for sufficiently large $k$.
Indeed, for any fixed integer $x\geq2$,
the corresponding sampling ratio
$\sigma=1/x$ gives
$\mathbb E[T_i]=k-r+r/x+O(k^{-1/x})$.
Consequently, for every fixed $r\geq1$,
\[
\lim_{k\to\infty}
\left(\mathbb E[T_i]\big|_{x=4}
-\mathbb E[T_i]\big|_{x=3}\right)
=-\frac{r}{12}<0.
\]
Thus $x=4$ eventually outperforms $x=3$,
including for simple parity codes ($r=1$).

More generally, for any fixed integers
$2\leq x<y$, the asymptotic difference
between their expected recovery times
is $r(1/y-1/x)<0$. Hence no fixed
multiplicity $x$ remains optimal as
$k\to\infty$.
\end{remark}

\subsection{Asymptotic Recovery Time for Linear Redundancy}
\label{subsec:mds-linear-r}

We now consider the regime where the redundancy grows
linearly with the dimension, i.e., $r/k\to\rho>0$.
Unlike the fixed-redundancy case, the expected recovery
time can be reduced by a constant fraction of $k$.

\begin{theorem}[Linear-Redundancy Asymptotics]
\label{thm:mds-linear-r}
Fix $\rho>0$ and $\sigma>0$. Let
$v_\rho(\sigma)\in(0,1)$ be the unique solution of
\[
v+\rho v^\sigma=\rho.
\]

For a systematic $[k+r,k]$ MDS code
with sampling ratio $\sigma$, if $r/k\to\rho$
as $k\to\infty$, then
\[
\lim_{k\to\infty}\frac{\mathbb E[T_i]}{k}
=(1+\rho\sigma)(1-v_\rho(\sigma)).
\]
\end{theorem}

\begin{proof}
For $u\in[0,1]$, let $A_u$ and $B_u$ be independent
random variables with
$A_u\sim\operatorname{Bin}(k-1,u)$ and
$B_u\sim\operatorname{Bin}(r,u^\sigma)$.
As in the proof of Theorem~\ref{thm:mds-fixed-r},
the exact recovery-time formula can be written as
\[
\frac{\mathbb E[T_i]}{k}
=\left(1+\frac rk\sigma\right)
\int_0^1\Pr(A_u+B_u\ge r)\,du.
\]

Set $X_{k,u}=A_u+B_u$ and
$g(u)=u+\rho u^\sigma$.
Since $r/k\to\rho$, we have
$\mathbb E[X_{k,u}]/k\to g(u)$.
Moreover,
$\operatorname{Var}(X_{k,u})
=(k-1)u(1-u)+ru^\sigma(1-u^\sigma)=O(k)$.

The function $g$ is continuous and strictly
increasing on $[0,1]$, with $g(0)=0<\rho$
and $g(1)=1+\rho>\rho$. Hence there is a
unique $v\in(0,1)$ satisfying $g(v)=\rho$.

Fix $u<v$. Since $g(u)<\rho$, there exists
$\delta>0$ such that, for all sufficiently
large $k$, we have
$r-\mathbb E[X_{k,u}]\ge\delta k$.
Chebyshev's inequality therefore gives
$\Pr(X_{k,u}\ge r)=O(k^{-1})$.

Similarly, for fixed $u>v$, we have
$\mathbb E[X_{k,u}]-r\ge\delta k$ for some
$\delta>0$ and all sufficiently large $k$.
Thus $\Pr(X_{k,u}<r)=O(k^{-1})$.

Consequently, for every $u\neq v$,
\[
\lim_{k\to\infty}\Pr(X_{k,u}\ge r)
=\begin{cases}
0,&u<v,\\
1,&u>v.
\end{cases}
\]
Since these probabilities are bounded by $1$,
the dominated convergence theorem gives
$\lim_{k\to\infty}\int_0^1
\Pr(X_{k,u}\ge r)\,du=1-v$.
The result follows from $r/k\to\rho$.
\end{proof}

%For fixed $\rho>0$, define
%$\Phi_\rho(\sigma)
%=(1+\rho\sigma)(1-v_\rho(\sigma))$.
%Under uniform sampling, $\sigma=1$, we have
%$v_\rho(1)=\rho/(1+\rho)$ and hence
%$\Phi_\rho(1)=1$, consistent with the
%exact recovery time $\mathbb E[T_i]=k$.
%
%The dependence of $\Phi_\rho$ on $\sigma$
%raises the natural question of which sampling
%ratio minimizes the asymptotic recovery time.
%We investigate this in the next subsection.

Under uniform sampling, $\sigma=1$, the defining
equation gives $v_\rho(1)=\rho/(1+\rho)$.
Consequently, the limiting normalized recovery
time equals $1$, consistent with the exact
identity $\mathbb E[T_i]=k$.

\subsection{Optimal Sampling Ratio}
\label{subsec:mds-optimal}

For fixed $\rho>0$, define
\[
\Phi_\rho(\sigma)
\triangleq (1+\rho\sigma)(1-v_\rho(\sigma)),
\text{ where }
v_\rho(\sigma)+\rho v_\rho(\sigma)^\sigma=\rho.
\]
The defining equation has a unique solution
$v_\rho(\sigma)\in(0,1)$. By
Theorem~\ref{thm:mds-linear-r},
$\Phi_\rho(\sigma)$ is the limiting normalized
expected recovery time.

\begin{theorem}[Optimal Sampling Ratio]
\label{thm:mds-optimal}
For every $\rho>0$, the function $\Phi_\rho$
has a unique global minimizer
$\sigma_\rho^\star\in(0,1)$.
The minimizer is characterized by
\[
\rho(1-v)[v+\sigma(\rho-v)]
=-(1+\rho\sigma)v(\rho-v)\ln v,
\]
where $v=v_\rho(\sigma)$ and
$\sigma=\sigma_\rho^\star$.
\end{theorem}

\begin{proof}
Fix $\rho>0$ and write $v=v_\rho(\sigma)$.
Implicit differentiation gives
\[
v'=-\frac{v(\rho-v)\ln v}{v+\sigma(\rho-v)}>0,
\qquad
\Phi_\rho'=\rho(1-v)-(1+\rho\sigma)v'.
\]
The stationary condition $\Phi_\rho'=0$
therefore gives the stated equation.

We first establish existence. From the defining
equation for $v$, we obtain
$\lim_{\sigma\to0^+}v=0$ and
$\lim_{\sigma\to1^-}v=\rho/(1+\rho)$.
Consequently, $\Phi_\rho$ tends to $1$
at both endpoints of $(0,1)$.

For $0<\sigma<1$, the strict concavity
of $v^\sigma$ gives
$v^\sigma<1-\sigma(1-v)$.
Together with $v=\rho(1-v^\sigma)$,
this implies $v>\rho\sigma(1-v)$,
and hence $\Phi_\rho(\sigma)<1$.
Thus $\Phi_\rho$ attains its minimum
at an interior point of $(0,1)$.
For $\sigma>1$, the corresponding
convexity inequality gives
$\Phi_\rho(\sigma)>1$, so this
minimum is also global over $\sigma>0$.

To establish uniqueness, we show in
Lemma~\ref{lem:mds-stationary-curvature}
of Appendix~\ref{app:mds-optimal} that
$\Phi_\rho''(\sigma)>0$ whenever
$\Phi_\rho'(\sigma)=0$.
Therefore every zero of $\Phi_\rho'$
is a crossing from negative to positive,
and there can be at most one such zero.
The interior minimizer is consequently
unique.
\end{proof}

\begin{table*}[!t]
\centering
\caption{Numerically optimized recovery times for systematic $[k+r,k]$ MDS codes. 
Each entry gives the minimum of $\mathbb E[T_i]/k$ 
with the corresponding optimal sampling ratio $\sigma_{k,r}^{\star}$ in parentheses.}
\label{tab:mds-numerical-optima}
\small
\setlength{\tabcolsep}{5pt}
\renewcommand{\arraystretch}{1.15}
\begin{tabular}{c|ccccc}
\hline
$k$ & $r=1$ & $r=2$ & $r=3$ & $r=4$ & $r=5$ \\
\hline
5 & 0.9595 (0.3935) & 0.9405 (0.3399) & 0.9291 (0.3020) & 0.9213 (0.2728) & 0.9156 (0.2492) \\
10 & 0.9717 (0.3757) & 0.9556 (0.3472) & 0.9444 (0.3263) & 0.9360 (0.3088) & 0.9294 (0.2934) \\
20 & 0.9823 (0.3574) & 0.9707 (0.3462) & 0.9618 (0.3379) & 0.9545 (0.3300) & 0.9484 (0.3221) \\
50 & 0.9914 (0.3337) & 0.9850 (0.3355) & 0.9795 (0.3368) & 0.9748 (0.3366) & 0.9706 (0.3355) \\
100 & 0.9952 (0.3169) & 0.9914 (0.3237) & 0.9880 (0.3285) & 0.9850 (0.3314) & 0.9822 (0.3331) \\
200 & 0.9974 (0.3011) & 0.9952 (0.3106) & 0.9933 (0.3172) & 0.9914 (0.3217) & 0.9897 (0.3249) \\
500 & 0.9989 (0.2823) & 0.9979 (0.2932) & 0.9970 (0.3006) & 0.9961 (0.3059) & 0.9953 (0.3099) \\
1000 & 0.9994 (0.2693) & 0.9989 (0.2804) & 0.9984 (0.2879) & 0.9979 (0.2933) & 0.9974 (0.2975) \\
\hline
\end{tabular}
\end{table*}

\begin{figure*}[!t]
\centering
\includegraphics[width=\textwidth]{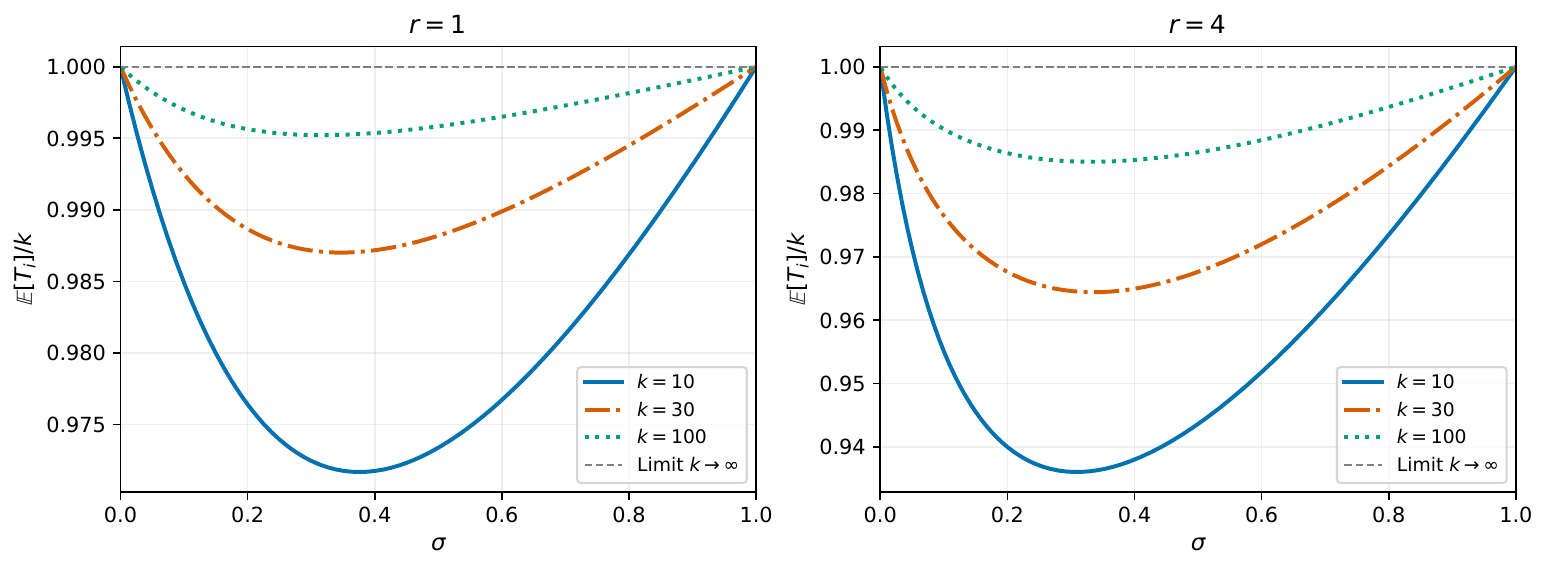}
\caption{Normalized exact recovery time for fixed redundancy
$r=1$ (left) and $r=4$ (right), with $k=10,30,100$.
The horizontal line at $1$ is the limiting normalized
recovery time for every fixed $0<\sigma\leq1$,
by Theorem~\ref{thm:mds-fixed-r} and the
uniform-sampling identity.}
\label{fig:mds-fixed-redundancy}
\end{figure*}

\begin{figure*}[!t]
\centering
\includegraphics[width=\textwidth]{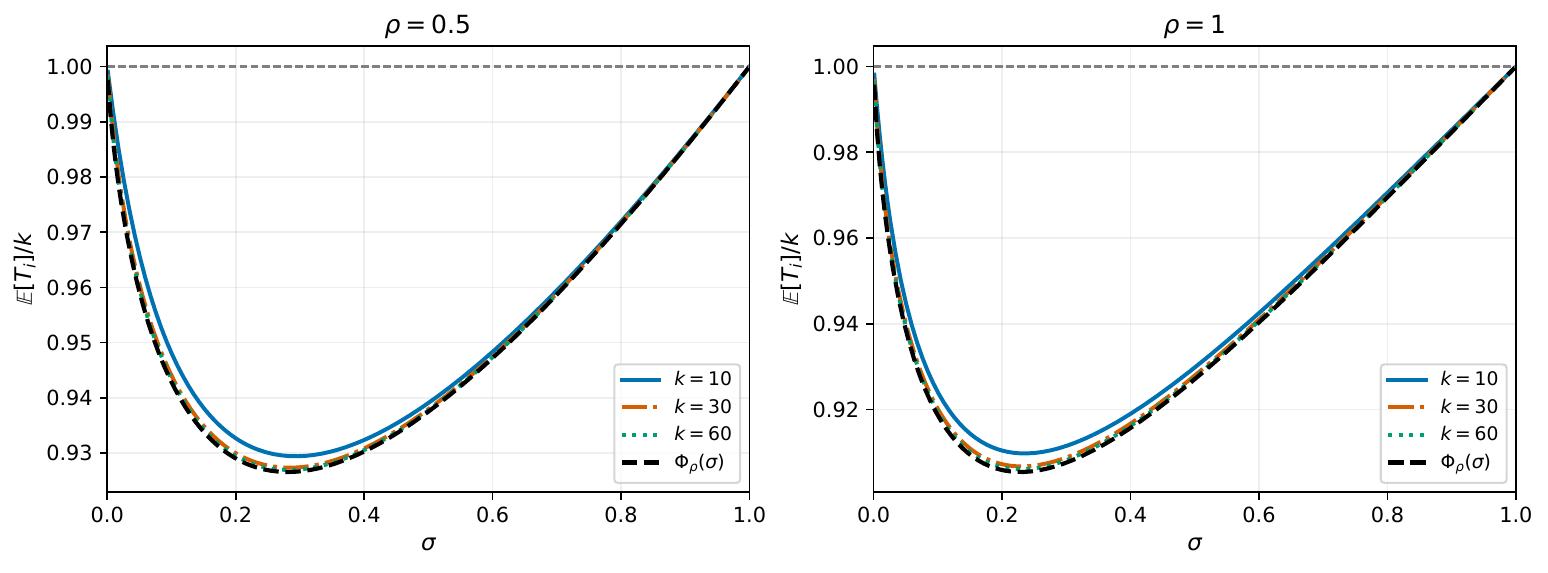}
\caption{Normalized exact recovery time for linear redundancy
$r=\rho k$, with $\rho=1/2$ (left), $\rho=1$ (right),
and $k=10,30,60$. The black dashed curve is
$\Phi_\rho(\sigma)$ from Theorem~\ref{thm:mds-linear-r}.}
\label{fig:mds-linear-redundancy}
\end{figure*}

\subsection{Numerical Illustrations}
\label{subsec:mds-numerical}

We numerically minimize the exact expectation in
Theorem~\ref{thm:mds-exact} over $\sigma>0$,
and denote a minimizing sampling ratio by
$\sigma_{k,r}^{\star}$. Table~\ref{tab:mds-numerical-optima} gives the
optimal normalized recovery times and their
corresponding sampling ratios.

Figure~\ref{fig:mds-fixed-redundancy} illustrates the fixed-redundancy regime.
For each fixed $r$, the normalized recovery time
approaches $1$ as $k$ increases, in agreement with
Theorem~\ref{thm:mds-fixed-r}.
In contrast, Figure~\ref{fig:mds-linear-redundancy} illustrates the
linear-redundancy regime, where the finite-length
curves approach the nontrivial limiting function
$\Phi_\rho(\sigma)$ of
Theorem~\ref{thm:mds-linear-r}.

\section{Recovery Time for Singleton--Pair Codes:
The Incidence Construction}
\label{sec:complete-graph}

We next apply the analytic combinatorial framework of Section~\ref{sec:ac-framework} to the singleton--pair construction. Its asymptotic recovery time under peeling decoding was derived by Hofmeister et al.~\cite[Section~V-D]{HofmeisterBitarYaakobi2026}, who expressed the limiting formula in terms of the dilogarithm and determined the optimal singleton mass. Bar-Lev et al.~\cite[Proposition~12]{BarLevElishcoGabrys2026} subsequently showed that the same leading limit holds under optimal decoding in their broader model over arbitrary fields and nonzero coefficient distributions. Both analyses use Poissonization, while the graph-component representation and Cayley's formula used below also appear in the work of Bar-Lev et al.

For the incidence construction, we evaluate the finite-length expectation through connectivity polynomials, obtain an explicit rational form, and derive the first finite-size correction, of order
$1/k$, to the limiting normalized recovery time.
These formulas also allow the sampling allocation to be optimized at finite blocklengths.

We consider the $[k+\binom{k}{2},k]$ code
with generator matrix $\bfG=(\bfI_k\mid\bfA)$,
where $\bfA$ is an oriented incidence matrix
of the complete graph $K_k$. Thus, the parity
columns are $\bfe_i-\bfe_j$, for
$1\le i<j\le k$.
The construction is defined over any field
$\mathbb F$. In particular, over a field of
characteristic two, the incidence columns
become $\bfe_i+\bfe_j$.

Each systematic column is sampled with probability
$p_{\mathrm{sys}}$ and each incidence column with
probability $p_{\mathrm{edge}}$, where
$kp_{\mathrm{sys}}+\binom{k}{2}p_{\mathrm{edge}}=1$.
Let $\alpha=kp_{\mathrm{sys}}$ denote the total
singleton sampling mass.
By symmetry, it suffices to study the recovery
time of $\mathbf e_1$, which we denote by $T_k$.

\subsection{Recovery Characterization}
\label{subsec:incidence-recovery}

Given a set $R$ of distinct sampled columns,
associate with $R$ a graph $H_R$ on vertex set $[k]$.
We include the edge $\{i,j\}$ in $H_R$ whenever
the incidence column $\bfe_i-\bfe_j$ belongs to $R$.
In addition, vertex $i$ is said to be sampled if
the systematic column $\bfe_i$ belongs to $R$.

The following lemma characterizes recovery entirely
in terms of the connected component containing the
requested vertex.

\begin{lemma}
\label{lem:incidence-recovery}
Let $S$ be the vertex set of the connected
component containing vertex $1$ in $H_R$.
Then $\bfe_1$ is recoverable from the columns
in $R$ if and only if $S$ contains a sampled vertex.
\end{lemma}

\begin{proof}
Suppose that some vertex $j\in S$ is sampled.
Since vertices $1$ and $j$ lie in the same
connected component, there exists a path
joining them. Adding the incidence columns
along this path, with appropriate signs,
gives $\bfe_1-\bfe_j$. Since $\bfe_j\in R$,
it follows that $\bfe_1$ lies in the span
of the sampled columns.

Conversely, suppose that $S$ contains no
sampled vertex. Every sampled column supported
on $S$ is then an incidence column
$\bfe_i-\bfe_j$ with $i,j\in S$.
Such columns have coordinate sum zero over $S$.
Moreover, no sampled incidence column joins
$S$ to its complement, and all remaining
sampled columns are supported outside $S$.
Consequently, every vector in the span of
$R$ has coordinate sum zero over $S$.
Since $\bfe_1$ has coordinate sum one over
$S$, it cannot lie in this span.
\end{proof}

\begin{remark}[Comparison with Related
Singleton--Pair Models]
In the binary degree-two construction of
Hofmeister et al.~\cite{HofmeisterBitarYaakobi2026},
the generator columns are $\bfe_i$ and
$\bfe_i+\bfe_j$, and recovery is performed
using a peeling decoder. Over $\mathbb F_2$,
these are precisely our columns, and peeling
recovers the vertices in components containing
a sampled singleton.

Bar-Lev et al.~\cite{BarLevElishcoGabrys2026} consider a more general singleton--pair model over arbitrary fields, with independently sampled nonzero coefficients on the two positions of each pair column, and analyze optimal decoding. 
With such coefficients, cycles in a component containing no sampled vertex can sometimes permit recovery. Nevertheless, their analysis shows that optimal and peeling decoding have the same asymptotic normalized recovery time for singleton--pair observations.

Boruchovsky et al.~\cite{Boruchovsky2026} studied recovery-complete constructions with multiple weight-two columns for each pair,
using coefficients over sufficiently large
finite fields. In those constructions, cycles
can also permit recovery without a sampled
systematic column.

For our incidence construction, however, every pair column has the form $\bfe_i-\bfe_j$.
The columns along each cycle are linearly dependent, over any field. 
Thus, a sampled singleton is necessary for recovery within the requested vertex's connected component,
as characterized in Lemma~\ref{lem:incidence-recovery}.
More importantly, this component criterion characterizes all non-recovery sets exactly,
for every blocklength and over any field.
This allows us to enumerate these sets through connected-graph polynomials and obtain
an explicit finite-length recovery-time formula.
\end{remark}

Lemma~\ref{lem:incidence-recovery} suggests that
non-recovery sets are easier to enumerate than
recovery sets. We therefore consider the
non-recovery EGF introduced in
Section~\ref{sec:ac-framework}.

For $s\ge1$, define the connected graph polynomial
$C_s(x)=\sum_H x^{|E(H)|}$, where the sum is over
all connected graphs $H$ on the labeled vertex
set $[s]$. In particular, $C_1(x)=1$.
Let $c_s(u)$ denote the probability that the
Erd\H{o}s--R\'enyi graph $G(s,u)$ is connected.
For $0\leq u<1$, these quantities satisfy
\[
c_s(u)=(1-u)^{\binom{s}{2}}
C_s\!\left(\frac{u}{1-u}\right).
\]

\begin{lemma}
\label{lem:incidence-nonrecovery-egf}
The non-recovery EGF of the incidence construction is
\[
\overline F_k(z)
=\sum_{s=1}^k\binom{k-1}{s-1}
C_s(e^{p_{\rm edge}z}-1)
e^{(k-s)p_{\rm sys}z+\binom{k-s}{2}p_{\rm edge}z}.
\]
\end{lemma}

\begin{proof}
Consider a non-recovery set $R$, and let
$S$ be the vertex set of the connected
component containing vertex $1$ in $H_R$.
Write $|S|=s$. There are
$\binom{k-1}{s-1}$ choices for $S$.
By Lemma~\ref{lem:incidence-recovery},
none of the systematic columns indexed by
$S$ may belong to $R$. Moreover, no
incidence column joining $S$ to its
complement may belong to $R$.

The sampled incidence columns within $S$
must form a connected graph. Since each
sampled incidence column contributes
$e^{p_{\rm edge}z}-1$, their combined
contribution is $C_s(e^{p_{\rm edge}z}-1)$.
All columns supported entirely outside
$S$ are unrestricted. There are $k-s$
systematic columns and $\binom{k-s}{2}$
incidence columns outside $S$, contributing
$e^{(k-s)p_{\rm sys}z}$ and
$e^{\binom{k-s}{2}p_{\rm edge}z}$,
respectively.
Multiplying these contributions and
summing over $s=1,\ldots,k$ gives the result.
\end{proof}

\subsection{Exact Finite-Length Recovery Time}
\label{subsec:incidence-exact}

We now apply the Laplace--Borel representation from
Theorem~\ref{thm:laplace-borel}.

\begin{proposition}
\label{prop:incidence-exact}
For the incidence construction described above,
\[
\mathbb E[T_k]
=\int_0^\infty\sum_{s=1}^k
\binom{k-1}{s-1}c_s(1-e^{-p_{\rm edge}t})
e^{-t[sp_{\rm sys}+s(k-s)p_{\rm edge}]}\,dt.
\]
\end{proposition}

\begin{proof}
By Theorem~\ref{thm:laplace-borel} and
Lemma~\ref{lem:incidence-nonrecovery-egf},
we have
$\mathbb E[T_k]=\int_0^\infty
e^{-t}\overline F_k(t)\,dt$.
Using $kp_{\rm sys}+\binom{k}{2}p_{\rm edge}=1$,
the contribution corresponding to a component
of size $s$ is
\[
\binom{k-1}{s-1}C_s(e^{p_{\rm edge}t}-1)
e^{-t[sp_{\rm sys}+s(k-s)p_{\rm edge}
+\binom{s}{2}p_{\rm edge}]}.
\]

Using the relation between $C_s$ and $c_s$ with
$u=1-e^{-p_{\rm edge}t}$ gives
\[
e^{-p_{\rm edge}t\binom{s}{2}}
C_s(e^{p_{\rm edge}t}-1)
=c_s(1-e^{-p_{\rm edge}t}).
\]
Substituting this identity into the
Laplace--Borel integral proves the result.
\end{proof}

The connectivity probabilities $c_s(u)$
can be computed recursively. We have
$c_1(u)=1$, and for $s\ge2$,
\[
c_s(u)=1-\sum_{j=1}^{s-1}
\binom{s-1}{j-1}c_j(u)(1-u)^{j(s-j)}.
\]
Indeed, if $G(s,u)$ is disconnected and
the component containing vertex $1$ has
size $j<s$, then its remaining $j-1$
vertices may be chosen in
$\binom{s-1}{j-1}$ ways. The induced graph
must be connected, while all $j(s-j)$
edges joining this component to its
complement must be absent. For example,
$c_2(u)=u$ and $c_3(u)=3u^2-2u^3$.

Furthermore, the integral in
Proposition~\ref{prop:incidence-exact}
can be evaluated term by term.

\begin{corollary}[Finite-Length Rational Form]
\label{cor:incidence-rational}
For $s\ge1$, write
$D_s(q)=c_s(1-q)
=\sum_{\ell=0}^{\binom{s}{2}}d_{s,\ell}q^\ell$.
Then
\[
\mathbb E[T_k]
=\sum_{s=1}^k\binom{k-1}{s-1}
\sum_{\ell=0}^{\binom{s}{2}}
\frac{d_{s,\ell}}
{sp_{\rm sys}+[s(k-s)+\ell]p_{\rm edge}}.
\]
In particular, $\mathbb E[T_k]$ is a
rational function of the sampling probabilities.
\end{corollary}

\begin{proof}
Substituting $u=1-q$ into the recursion
for $c_s(u)$ gives $D_1(q)=1$ and
\[
D_s(q)=1-\sum_{h=1}^{s-1}
\binom{s-1}{h-1}D_h(q)q^{h(s-h)}.
\]
Comparing coefficients yields
\[
d_{s,\ell}
=\mathbf 1_{\{\ell=0\}}
-\sum_{h=1}^{s-1}
\binom{s-1}{h-1}d_{h,\ell-h(s-h)},
\]
where $d_{h,j}=0$ for $j<0$ or
$j>\binom{h}{2}$.

Now substitute
$c_s(1-e^{-p_{\rm edge}t})
=D_s(e^{-p_{\rm edge}t})$
into Proposition~\ref{prop:incidence-exact}.
Expanding the polynomial produces terms
of the form
$e^{-t[sp_{\rm sys}+
(s(k-s)+\ell)p_{\rm edge}]}$.
Integrating these terms over
$t\in[0,\infty)$ gives the result.
\end{proof}

Thus, for $p_{\rm sys}>0$, the connectivity
recursion gives an exact rational expression
for $\mathbb E[T_k]$. In particular, it permits
numerical optimization of finite-length sampling
allocations without numerical integration.
For the asymptotic analysis, we return to
the integral representation.

\subsection{Second-Order Approximation of Recovery Time}
\label{subsec:incidence-second-order}

We now derive a second-order approximation to
the normalized recovery time
$\Phi_k(\alpha)\triangleq\mathbb E[T_k]/k$.
More precisely, we identify the first
finite-length correction, which appears at order
$1/k$, while the next correction term is
$O_\alpha(k^{-2})$.
The proof below derives the correction term, while
the technical estimates needed to justify
the stated error bound are deferred to Appendix~\ref{app:second-order-error}.

To describe the correction term, we recall two
enumerative results concerning connected graphs.
By Cayley's formula, there are $s^{s-2}$ connected
labeled graphs on $s$ vertices with $s-1$ edges.
Let $a_s$ denote the number of connected labeled
graphs on $s$ vertices with $s$ edges, or equivalently,
the number of labeled unicyclic graphs
(OEIS sequence A057500).
Consequently, the connected-graph polynomial
has the expansion
$C_s(z)=s^{s-2}z^{s-1}+a_s z^s+O(z^{s+1})$.

\begin{theorem}[Second-Order Approximation]
\label{thm:incidence-second-order}
Fix $0<\alpha<1$ and set
$x=2(1-\alpha)/(2-\alpha)$. Then
\[
\Phi_k(\alpha)
=
\Phi(\alpha)
+\frac{A(\alpha)}{k-1}
+O_\alpha(k^{-2}),
\]
where
\[
\Phi(\alpha)
=
\frac{2-x}{2x}\operatorname{Li}_2(x)
=
\frac{1}{2(1-\alpha)}
\operatorname{Li}_2\left(
\frac{2(1-\alpha)}{2-\alpha}\right),
\]
and
\[
A(\alpha)
=
\frac{2-x}{2}
\sum_{s\ge1}
\frac{x^{s-1}}{s^2}
\left[
-\frac{(s-1)(s-2)}{2}
+
x\left(
\frac{a_s}{s^{s-2}}
+
\frac{(s-1)(s+1)}{2}
\right)
\right].
\]
\end{theorem}

\begin{proof}
Starting from Proposition~\ref{prop:incidence-exact}, substituting \(p_{\rm sys}=\alpha/k\), \(p_{\rm edge}=2(1-\alpha)/(k(k-1))\), and setting \(t=kr\), we obtain
\[
\Phi_k(\alpha)
=
\sum_{s=1}^k
\binom{k-1}{s-1}
\int_0^\infty
e^{-(2-\alpha)sr}
e^{\frac{(1-\alpha)s(s-1)r}{k-1}}
C_s\!\left(e^{\frac{2(1-\alpha)r}{k-1}}-1\right)\,dr.
\]

By Cayley's formula, \([z^{s-1}]C_s(z)=s^{s-2}\). Let \(a_s=[z^s]C_s(z)\), and write
\(C_s(z)=s^{s-2}z^{s-1}+a_sz^s+R_s(z)\).
Substituting this expansion into the preceding expression, let \(\Phi_k^{(1)}\), \(\Phi_k^{(2)}\), and \(\Phi_k^{(\mathrm{rem})}\) denote the contributions obtained by replacing \(C_s(z)\) with \(s^{s-2}z^{s-1}\), \(a_sz^s\), and \(R_s(z)\), respectively. Thus
$\Phi_k=\Phi_k^{(1)}+\Phi_k^{(2)}+\Phi_k^{(\mathrm{rem})}$.

For \(\Phi_k^{(1)}\), we expand the three \(k\)-dependent factors to first order in \((k-1)^{-1}\):
\[
\begin{aligned}
\binom{k-1}{s-1}
&=
\frac{(k-1)^{s-1}}{(s-1)!}
\left(
1-\frac{(s-1)(s-2)}{2(k-1)}
+O_\alpha(k^{-2})
\right),\\
e^{\frac{(1-\alpha)s(s-1)r}{k-1}}
&=
1+\frac{(1-\alpha)s(s-1)r}{k-1}
+O_\alpha(k^{-2}),\\
\left(e^{\frac{2(1-\alpha)r}{k-1}}-1\right)^{s-1}
&=
\left(\frac{2(1-\alpha)r}{k-1}\right)^{s-1}
\left(
1+\frac{(s-1)(1-\alpha)r}{k-1}
+O_\alpha(k^{-2})
\right).
\end{aligned}
\]
Multiplying these expansions gives
\[
\begin{aligned}
\Phi_k^{(1)}(\alpha)
={}&
\sum_{s\ge1}
\frac{s^{s-2}[2(1-\alpha)]^{s-1}}{(s-1)!}
\int_0^\infty r^{s-1}e^{-(2-\alpha)sr}
\left[
1+\frac{1}{k-1}
\left(
-\frac{(s-1)(s-2)}2
+(1-\alpha)(s-1)(s+1)r
\right)
\right]dr \\
& \qquad + O_\alpha(k^{-2}).
\end{aligned}
\]

The uniform error bounds justifying these expansions and the interchange of the summation and integration are established in Proposition~\ref{prop:second-order-error} of Appendix~\ref{app:second-order-error}.

Now set \(x=2(1-\alpha)/(2-\alpha)\). Using
\[
\int_0^\infty r^{s-1}e^{-(2-\alpha)sr}\,dr
=
\frac{(s-1)!}{[(2-\alpha)s]^s},
\text{ and }
\int_0^\infty r^{s}e^{-(2-\alpha)sr}\,dr
=
\frac{s!}{[(2-\alpha)s]^{s+1}},
\]
we obtain
\[
\Phi_k^{(1)}(\alpha)
=
\Phi(\alpha)
+
\frac{1}{k-1}\frac{2-x}{2}
\sum_{s\ge1}\frac{x^{s-1}}{s^2}
\left[
-\frac{(s-1)(s-2)}2
+
\frac{x(s-1)(s+1)}2
\right]
+
O_\alpha(k^{-2}),
\]
where
\(\Phi(\alpha)=\frac{2-x}{2x}\operatorname{Li}_2(x)\).

The term \(\Phi_k^{(2)}\) already carries one additional factor of \((k-1)^{-1}\), so only the leading terms of the remaining factors are required. This gives
\[
\Phi_k^{(2)}(\alpha)
=
\frac{1}{k-1}\frac{2-x}{2}
\sum_{s\ge1}\frac{x^{s-1}}{s^2}
\left(
x\frac{a_s}{s^{s-2}}
\right)
+
O_\alpha(k^{-2}).
\]

Finally, we prove \(\Phi_k^{(\mathrm{rem})}(\alpha)=O_\alpha(k^{-2})\) in Appendix~\ref{app:second-order-error}. Combining the three terms corresponding to $(k-1)^{-1}$, we have
\[
A(\alpha)
=
\frac{2-x}{2}
\sum_{s\ge1}
\frac{x^{s-1}}{s^2}
\left[
-\frac{(s-1)(s-2)}2
+
x\left(
\frac{a_s}{s^{s-2}}
+
\frac{(s-1)(s+1)}2
\right)
\right].
\]
\end{proof}

The leading term $\Phi(\alpha)$ is the
singleton--pair peeling asymptotic derived and
optimized by Hofmeister et al.~\cite[Section~V-D]{HofmeisterBitarYaakobi2026}.
Bar-Lev et al.~\cite[Proposition~12]{BarLevElishcoGabrys2026}
established the same leading limit under optimal
decoding in their broader field and coefficient model.
The unique minimizer is
$\alpha^\star\approx0.1554737004$, with
$\Phi(\alpha^\star)\approx0.7939334446$.
Since incidence recovery has the same component
recovery criterion as binary peeling, Lemma~2 and Theorem~2
of~\cite{HofmeisterBitarYaakobi2026} also imply that
the exact finite-length optimizers converge to
$\alpha^\star$.

The second-order expansion suggests the
finite-length approximation
$\widetilde{\Phi}_k(\alpha)
\triangleq\Phi(\alpha)+A(\alpha)/(k-1)$.
Let $\widetilde{\alpha}_k$ minimize
$\widetilde{\Phi}_k$, and let
$\alpha_k^\star$ minimize the exact
finite-length quantity $\Phi_k$.
Unlike the exact rational expression,
$\widetilde{\Phi}_k$ requires only the
evaluation of a convergent series and can
therefore be optimized without computing all
connectivity polynomials up to size $k$.

\begin{table*}[!t]
\centering
\caption{Exact and second-order optimized recovery
times. Here $\alpha_k^\star$ minimizes the exact
finite-length quantity $\Phi_k$, while
$\widetilde{\alpha}_k$ minimizes the second-order
approximation $\widetilde{\Phi}_k$.}
\label{tab:second-order-optimal}
\begin{tabular}{|c|c|c|c|c|c|}
\hline
$k$
&
$\alpha_k^\star$
&
$\widetilde{\alpha}_k$
&
$\Phi_k(\alpha_k^\star)$
&
$\widetilde{\Phi}_k(\widetilde{\alpha}_k)$
&
$\Phi_k(\alpha_k^\star)
-\widetilde{\Phi}_k(\widetilde{\alpha}_k)$
\\
\hline
20 & 0.352726 & 0.320773 & 0.835523 & 0.831707 & 0.003816 \\
30 & 0.303899 & 0.280527 & 0.824108 & 0.821525 & 0.002583 \\
40 & 0.275280 & 0.257255 & 0.817691 & 0.815811 & 0.001880 \\
50 & 0.256225 & 0.241806 & 0.813554 & 0.812117 & 0.001437 \\
60 & 0.242535 & 0.230690 & 0.810656 & 0.809518 & 0.001138 \\
\hline
\end{tabular}
\end{table*}

\begin{figure*}[!t]
\centering
\begin{tabular}{@{}c@{\hspace{0.02\textwidth}}c@{}}
\includegraphics[width=0.49\textwidth]{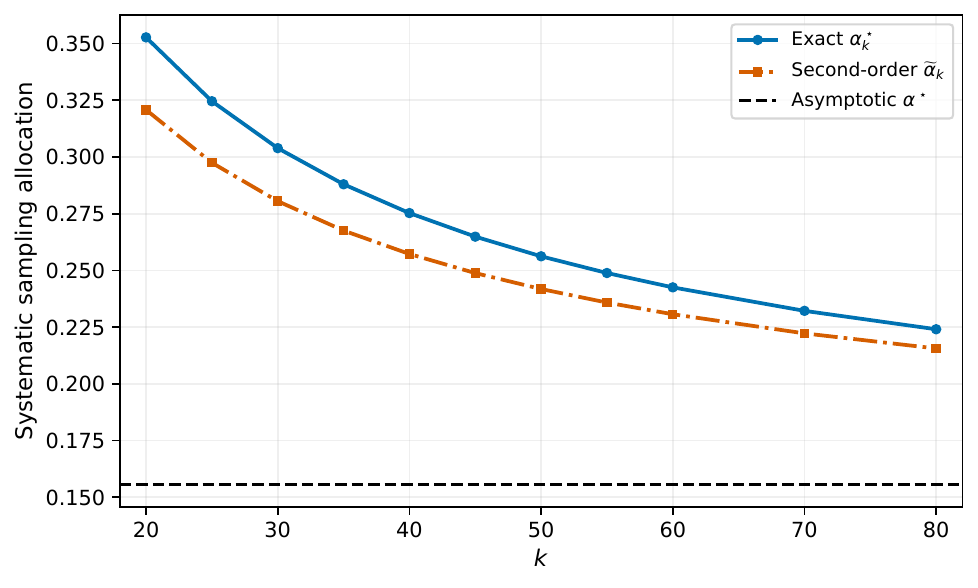}
&
\includegraphics[width=0.49\textwidth]{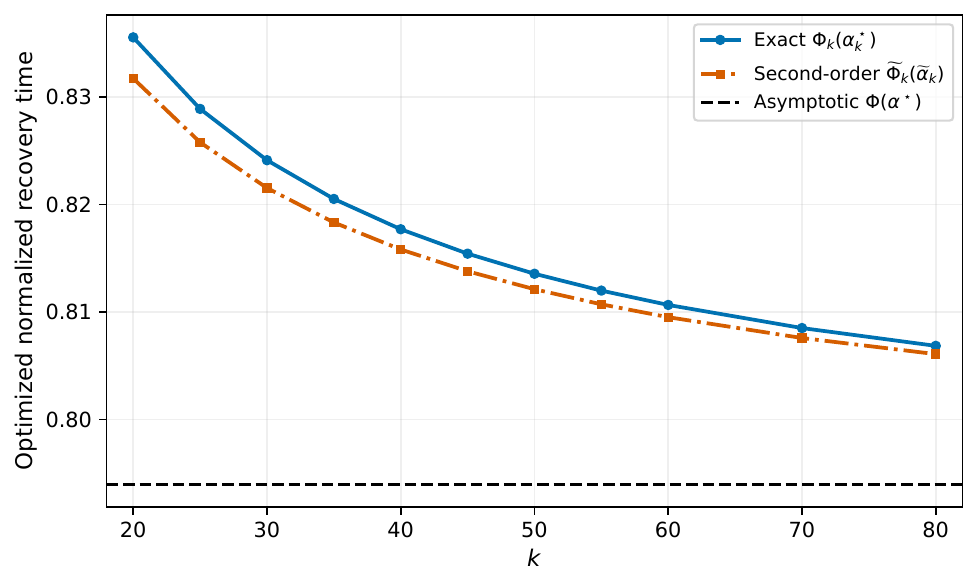}
\end{tabular}
\caption{Exact and second-order optimal systematic sampling
allocations (left) and optimized recovery times (right) versus
blocklength $k$. The horizontal dashed lines show the asymptotic
optima $\alpha^\star$ (left) and $\Phi(\alpha^\star)$ (right).}
\label{fig:finite_length_convergence}
\end{figure*}

Figure~\ref{fig:finite_length_convergence} compares
the optimizing allocations (left) and the corresponding
optimized recovery times (right) across blocklengths.
The second-order approximation captures most of
the finite-length displacement from the
asymptotic optimum. For example, at $k=60$,
the asymptotic optimizer is
$\alpha^\star\approx0.155474$, whereas the
second-order approximation gives
$\widetilde{\alpha}_{60}\approx0.230690$,
compared with the exact value
$\alpha_{60}^\star\approx0.242535$.
Likewise,
$\widetilde{\Phi}_{60}
(\widetilde{\alpha}_{60})\approx0.809518$,
compared with the exact normalized recovery time
$\Phi_{60}(\alpha_{60}^\star)\approx0.810656$.
Figure~\ref{fig:finite_length_convergence} shows that
the optimizing allocation (left) converges rather slowly to
$\alpha^\star$, whereas the optimized recovery time (right)
is already much closer to its asymptotic limit.
Thus the second-order term explains much of the
slow finite-length convergence while remaining
substantially simpler to evaluate than the exact
finite-length formula.

\section*{Acknowledgment}

We thank Eitan Yaakobi and Avital Boruchovsky for early discussions on the random-access problem, particularly on the expected recovery time of single-parity-check codes.

We also acknowledge the use of OpenAI's ChatGPT and Codex for assistance with mathematical exploration, symbolic and numerical computations, and manuscript preparation. 
The authors take full responsibility for the correctness of all results and arguments presented in this paper.

\appendices

\section{Uniqueness of the Optimal Sampling Ratio}
\label{app:mds-optimal}

We establish the positivity of the second derivative
at every stationary point of $\Phi_\rho$.

\begin{lemma}
\label{lem:mds-stationary-curvature}
Fix $\rho>0$. For every $\sigma\in(0,1)$
satisfying $\Phi_\rho'(\sigma)=0$, we have
$\Phi_\rho''(\sigma)>0$.
\end{lemma}

\begin{proof}
Write $v=v_\rho(\sigma)$ and set
$L=-\ln v>0$, $A=\rho-v$, and
$D=v+\sigma(\rho-v)$.
Implicit differentiation gives $v'=vAL/D$.
Taking a logarithmic derivative, using
$A'=-v'$ and $D'=A+(1-\sigma)v'$,
we obtain
\[
\frac{v''}{(v')^2}
=\frac1v-\frac1A-\frac2{vL}
-\frac{1-\sigma}{D}.
\]

At a stationary point,
$\rho(1-v)=(1+\rho\sigma)v'$.
Substituting into
$\Phi_\rho''=-2\rho v'-(1+\rho\sigma)v''$
and using $1/A+(1-\sigma)/D=\rho/(AD)$,
we obtain
\[
\Phi_\rho''
=(1+\rho\sigma)(v')^2
\left[
\frac{\rho}{AD}+\frac2{vL}
-\frac1v-\frac2{1-v}
\right].
\]

Set $w=v^\sigma=(\rho-v)/\rho$
and $B=1-w+\sigma w$.
Since $\rho=v/(1-w)$ and
$D=vB/(1-w)$, this simplifies to
\[
\Phi_\rho''
=\frac{(1+\rho\sigma)(v')^2}{v}
\left[
\frac{1-w}{wB}+\frac2L
-\frac{1+v}{1-v}
\right].
\]
It remains to prove that the quantity
in brackets is positive.

First, we claim that
$(1-w)/(wB)>L/(1+L)$.
Let $m=-\ln w=\sigma L$, so
$0<m<L$ and $w=e^{-m}$. Then
\[
\begin{aligned}
(1+L)(1-w)-LwB
&=(1-w)+L(1-w)^2-mw^2\\
&>1+m-(1+2m)e^{-m}>0.
\end{aligned}
\]
The last inequality follows from
$e^m>1+m$ and $(1+m)^2>1+2m$.

Second, since $v=e^{-L}$ and
$e^L-1>L+L^2/2$, we have
\[
\begin{aligned}
\frac{L}{1+L}+\frac2L-\frac{1+v}{1-v}
&=\frac{L+2}{L(1+L)}-\frac2{e^L-1}\\
&>\frac{L}{(1+L)(L+2)}>0.
\end{aligned}
\]

Combining these inequalities gives
\[
\frac{1-w}{wB}+\frac2L-\frac{1+v}{1-v}
>\frac{L}{(1+L)(L+2)}>0.
\]
Since the prefactor in the expression
for $\Phi_\rho''$ is strictly positive,
the result follows.
\end{proof}

\section{Error Bounds for Theorem~\ref{thm:incidence-second-order}}
\label{app:second-order-error}

In this appendix, we justify the error bounds used in the proof of
Theorem~\ref{thm:incidence-second-order}. Specifically, we prove the
following proposition. Recall that
\(x=2(1-\alpha)/(2-\alpha)\),
\(a_s=[z^s]C_s(z)\), and that
\(\Phi_k^{(1)}\), \(\Phi_k^{(2)}\), and
\(\Phi_k^{(\mathrm{rem})}\) are the contributions corresponding to
\(s^{s-2}z^{s-1}\), \(a_sz^s\), and \(R_s(z)\), respectively.

\begin{proposition}
\label{prop:second-order-error}
For fixed \(0<\alpha<1\),
\[
\begin{aligned}
\text{\rm (i)}\quad
\Phi_k^{(1)}(\alpha)
={}&
\Phi(\alpha)
+\frac{1}{k-1}\frac{2-x}{2}
\sum_{s\ge1}\frac{x^{s-1}}{s^2}
\left[
-\frac{(s-1)(s-2)}2
+\frac{x(s-1)(s+1)}2
\right]
+O_\alpha(k^{-2}),\\
\text{\rm (ii)}\quad
\Phi_k^{(2)}(\alpha)
={}&
\frac{1}{k-1}\frac{2-x}{2}
\sum_{s\ge1}\frac{x^{s-1}}{s^2}
\left(
x\frac{a_s}{s^{s-2}}
\right)
+O_\alpha(k^{-2}),\\
\text{\rm (iii)}\quad
\Phi_k^{(\mathrm{rem})}(\alpha)
={}&O_\alpha(k^{-2}).
\end{aligned}
\]
\end{proposition}

\begin{proof}
We first restrict the range of \(s\) and \(r\) on which the expansions will be carried out. We show that there exist constants \(C_1,C_2>0\), depending only on the fixed \(\alpha\), such that the total contribution from \(s>C_1\ln k\) or \(r>C_2\ln k\) is \(O_\alpha(k^{-3})\).

For each \(r\ge0\), let
$H_r\sim G\!\left(k,1-e^{-\frac{2(1-\alpha)r}{k-1}}\right)$,
the Erd\H{o}s--R\'enyi graph on $k$ vertices with edges chosen with probability $1-e^{-\frac{2(1-\alpha)r}{k-1}}$.
Let \(S_r\) be the vertex set of the connected component containing vertex \(1\) in \(H_r\). 
Using 
$C_s\!\left(e^{\frac{2(1-\alpha)r}{k-1}}-1\right)
=
e^{\frac{(1-\alpha)s(s-1)r}{k-1}}
c_s\!\left(1-e^{-\frac{2(1-\alpha)r}{k-1}}\right)$, we have
\[
\Phi_k(\alpha)
=
\int_0^\infty
\sum_{s=1}^k
e^{-\alpha sr}\Pr\{|S_r|=s\}\,dr,
\]
where
\[
\Pr\{|S_r|=s\}
=
\binom{k-1}{s-1}
e^{-\frac{2(1-\alpha)s(k-s)r}{k-1}}
c_s\!\left(
1-e^{-\frac{2(1-\alpha)r}{k-1}}
\right).
\]

We first control the contribution from large \(r\). Since the probabilities \(\Pr\{|S_r|=s\}\) sum to one,
\[
\int_{C_2\ln k}^{\infty}
\sum_{s=1}^k
e^{-\alpha sr}\Pr\{|S_r|=s\}\,dr
\le
\int_{C_2\ln k}^{\infty}e^{-\alpha r}\,dr
=
\frac{k^{-\alpha C_2}}{\alpha}.
\]
Thus this contribution is \(O_\alpha(k^{-3})\) provided \(C_2>3/\alpha\).

It remains to control large \(s\) for \(0\le r\le C_2\ln k\). Set $r_0=\frac{1}{4e(1-\alpha)}$.
The reason for this choice will appear below. 
For \(r\ge r_0\),
\[
\int_{r_0}^{C_2\ln k}
\sum_{s>C_1\ln k}
e^{-\alpha sr}\Pr\{|S_r|=s\}\,dr
\le
\int_{r_0}^{\infty}
e^{-\alpha C_1r\ln k}\,dr
=
\frac{k^{-\alpha C_1r_0}}
{\alpha C_1\ln k}.
\]
Hence this contribution is \(O_\alpha(k^{-3})\) if \(C_1\) is chosen sufficiently large.

Finally, consider \(0\le r\le r_0\). If \(|S_r|=s\), then the subgraph induced by \(S_r\) is connected and therefore contains a spanning tree. A union bound over the choice of the \(s-1\) vertices other than vertex \(1\), and over the \(s^{s-2}\) labeled spanning trees on these \(s\) vertices, gives
\[
\Pr\{|S_r|=s\}
\le
\binom{k-1}{s-1}s^{s-2}
\left(
1-e^{-\frac{2(1-\alpha)r}{k-1}}
\right)^{s-1}
\le
\frac{s^{s-2}[2(1-\alpha)r]^{s-1}}{(s-1)!}.
\]
Using \((s-1)!\ge((s-1)/e)^{s-1}\), the right-hand side is at most
\(e[2e(1-\alpha)r]^{s-1}\). By the choice of \(r_0\),
\(2e(1-\alpha)r\le1/2\) throughout this range, and hence
\[
\Pr\{|S_r|=s\}\le e\,2^{-(s-1)}.
\]
Consequently,
\[
\int_0^{r_0}
\sum_{s>C_1\ln k}
e^{-\alpha sr}\Pr\{|S_r|=s\}\,dr
\le
e r_0\sum_{s>C_1\ln k}2^{-(s-1)}
=
O\!\left(k^{-C_1\ln2}\right).
\]
After increasing \(C_1\), if necessary, this is \(O_\alpha(k^{-3})\).

We may therefore choose \(C_1,C_2>0\), depending only on \(\alpha\), such that the total contribution from
\(s>C_1\ln k\) or \(r>C_2\ln k\) is \(O_\alpha(k^{-3})\).
Since the three terms in
\(C_s(z)=s^{s-2}z^{s-1}+a_sz^s+R_s(z)\)
have nonnegative coefficients for \(z\ge0\), the same estimate applies separately to
\(\Phi_k^{(1)}\), \(\Phi_k^{(2)}\), and
\(\Phi_k^{(\mathrm{rem})}\).

On this truncated range,
\[
\frac{s^2r}{k-1}
\le
C_1^2C_2\frac{(\ln k)^3}{k-1}
=o(1),
\]
so the three expansions used in the proof of
Theorem~\ref{thm:incidence-second-order} hold with error bounds
uniform in \(s\) and \(r\). More precisely,
\[
\begin{aligned}
\binom{k-1}{s-1}
&=
\frac{(k-1)^{s-1}}{(s-1)!}
\left[
1-\frac{(s-1)(s-2)}{2(k-1)}
+O\!\left(\frac{s^4}{(k-1)^2}\right)
\right],\\
e^{\frac{(1-\alpha)s(s-1)r}{k-1}}
&=
1+\frac{(1-\alpha)s(s-1)r}{k-1}
+O_\alpha\!\left(\frac{s^4r^2}{(k-1)^2}\right),\\
\left(e^{\frac{2(1-\alpha)r}{k-1}}-1\right)^{s-1}
&=
\left(\frac{2(1-\alpha)r}{k-1}\right)^{s-1}
\left[
1+\frac{(s-1)(1-\alpha)r}{k-1}
+O_\alpha\!\left(\frac{s^2r^2}{(k-1)^2}\right)
\right].
\end{aligned}
\]
Multiplying the three expansions, the first-order terms are
\[
-\frac{(s-1)(s-2)}{2}
+(1-\alpha)(s-1)(s+1)r,
\]
while the second-order remainders and the products of first-order
terms are together
\[
O_\alpha\!\left(
\frac{(1+s)^4(1+r)^2}{(k-1)^2}
\right).
\]
Consequently, the difference between the \(s\)-th integrand of
\(\Phi_k^{(1)}\) and the first-order expression used in the main proof
is bounded by
\[
\frac{C_\alpha}{(k-1)^2}
\frac{s^{s-2}[2(1-\alpha)]^{s-1}}{(s-1)!}
r^{s-1}e^{-(2-\alpha)sr}
(1+s)^4(1+r)^2.
\]
By increasing \(C_1\) and \(C_2\), if necessary, the corresponding tails of the approximating expressions are also \(O_\alpha(k^{-3})\). Indeed, after integration in \(r\), their \(s\)-th terms are bounded by a polynomial in \(s\) times \(x^{s-1}\), so the contribution from \(s>C_1\ln k\) is \(O_\alpha(k^{-3})\) for sufficiently large \(C_1\). For the \(r\)-tail, multiply the integrands by \(e^{\alpha r/2}\). The same Gamma-integral calculation then gives a convergent series in \(s\), since
$
\frac{2(1-\alpha)}{2-\frac{3\alpha}{2}}<1.
$
It follows that the contribution from \(r>C_2\ln k\) is
\(O_\alpha(k^{-\alpha C_2/2})\), and hence \(O_\alpha(k^{-3})\) for sufficiently large \(C_2\).

We may therefore enlarge the truncated integral to \(r\in[0,\infty)\) and the
truncated sum to \(s\ge1\). For \(j=0,1,2\),
\[
\int_0^\infty
r^{s-1+j}e^{-(2-\alpha)sr}\,dr
=
\frac{(s+j-1)!}{[(2-\alpha)s]^{s+j}}.
\]
Hence, after integration, the error is bounded by
\[
\frac{C_\alpha}{(k-1)^2}
\sum_{s\ge1}
\frac{x^{s-1}}{s^2}(1+s)^4
\left[
1+
\frac{2s}{(2-\alpha)s}
+
\frac{s(s+1)}{[(2-\alpha)s]^2}
\right].
\]
The expression in square brackets is bounded uniformly in \(s\), and
the remaining series converges since \(0<x<1\). Thus the accumulated
error is \(O_\alpha(k^{-2})\), proving part~{\rm (i)}.

We proceed similarly for \(\Phi_k^{(2)}\). Since this term contains one additional power of
\(e^{2(1-\alpha)r/(k-1)}-1\), it already carries a factor
\((k-1)^{-1}\). On the truncated range, the same estimates as above give
\[
\begin{aligned}
&\frac{\binom{k-1}{s-1}(s-1)!}{(k-1)^{s-1}}
e^{\frac{(1-\alpha)s(s-1)r}{k-1}}
\left[
\frac{e^{2(1-\alpha)r/(k-1)}-1}
     {2(1-\alpha)r/(k-1)}
\right]^s\\
&\qquad =
1+
O_\alpha\!\left(
\frac{(1+s)^2(1+r)}{k-1}
\right).
\end{aligned}
\]
Here the quotient is interpreted by continuity at \(r=0\).
Hence
\[
\Phi_k^{(2)}(\alpha)
=
\frac{1}{k-1}
\sum_{s\ge1}
\frac{a_s[2(1-\alpha)]^s}{(s-1)!}
\int_0^\infty
r^s e^{-(2-\alpha)sr}\,dr
+
O_\alpha(k^{-2}),
\]
provided that the accumulated error is summable. To verify this, note that a connected graph on \(s\) vertices with \(s\) edges contains a spanning tree together with one additional edge, and therefore
\(a_s\le s^{s-2}\binom{s}{2}\).
The error is consequently bounded by \(C_\alpha(k-1)^{-2}\) times a sum of Gamma moments whose \(s\)-th term is at most a polynomial in \(s\) times \(x^s\). Since \(0<x<1\), this sum converges. Evaluating the displayed integral gives
\[
\Phi_k^{(2)}(\alpha)
=
\frac{1}{k-1}\frac{2-x}{2}
\sum_{s\ge1}
\frac{x^{s-1}}{s^2}
\left(
x\frac{a_s}{s^{s-2}}
\right)
+
O_\alpha(k^{-2}),
\]
which proves part~{\rm (ii)}.

Finally, we consider \(\Phi_k^{(\mathrm{rem})}\). Recall that
\[
R_s(z)
=
\sum_{j\ge2}[z^{s-1+j}]C_s(z)\,z^{s-1+j}.
\]
Every connected graph counted by \([z^{s-1+j}]C_s(z)\) contains a spanning tree. A union bound over the spanning tree and the \(j\) additional edges gives
\[
[z^{s-1+j}]C_s(z)
\le
s^{s-2}\binom{\binom{s}{2}}{j}.
\]
Thus, for \(z\ge0\),
\[
R_s(z)
\le
s^{s-2}z^{s-1}
\sum_{j\ge2}\frac{(s^2z)^j}{j!}
\le
s^{s+2}z^{s+1}e^{s^2z}.
\]

On the truncated range, let
\(z=e^{2(1-\alpha)r/(k-1)}-1\).
Since
\[
z\le
\frac{2(1-\alpha)r}{k-1}
e^{\frac{2(1-\alpha)r}{k-1}},
\]
we have \(s^2z=o(1)\) uniformly. Moreover,
\(s^2r/(k-1)=o(1)\), so the remaining exponential factor
\(e^{(1-\alpha)s(s-1)r/(k-1)}\) is uniformly bounded. It follows that
\[
R_s(z)
\le
C_\alpha s^{s+2}
\left(
\frac{2(1-\alpha)r}{k-1}
\right)^{s+1}
\]
throughout the truncated range.

Using
\(\binom{k-1}{s-1}\le (k-1)^{s-1}/(s-1)!\),
the remainder contribution is therefore bounded by
\[
\frac{C_\alpha}{(k-1)^2}
\sum_{s\ge1}
\frac{s^{s+2}[2(1-\alpha)]^{s+1}}{(s-1)!}
\int_0^\infty
r^{s+1}e^{-(2-\alpha)sr}\,dr.
\]
The Gamma integral equals
\((s+1)!/[(2-\alpha)s]^{s+2}\), and hence the last expression is
\[
\frac{C_\alpha}{(k-1)^2}
\sum_{s\ge1}s(s+1)x^{s+1}.
\]
This series converges because \(0<x<1\). Together with the
\(O_\alpha(k^{-3})\) contribution discarded in the truncation step, we obtain
\(\Phi_k^{(\mathrm{rem})}(\alpha)=O_\alpha(k^{-2})\), proving part~{\rm (iii)}.
\end{proof}

%\newpage
%\input{MDS_Section-OLD}
%\newpage
%\input{Complete_Graph_Incidence_Codes}

\bibliographystyle{IEEEtran}
%\newpage
\bibliography{references}

@article{FlajoletGardyThimonier1992,
  author  = {Philippe Flajolet and Dani{\`e}le Gardy and Lo{\"y}s Thimonier},
  title   = {Birthday Paradox, Coupon Collectors, Caching Algorithms and Self-Organizing Search},
  journal = {Discrete Applied Mathematics},
  volume  = {39},
  number  = {3},
  pages   = {207--229},
  year    = {1992},
  doi     = {10.1016/0166-218X(92)90177-C}
}

@book{FlajoletSedgewick2009,
  author    = {Philippe Flajolet and Robert Sedgewick},
  title     = {Analytic Combinatorics},
  publisher = {Cambridge University Press},
  address   = {Cambridge, UK},
  year      = {2009},
  isbn      = {9780521898065},
  doi       = {10.1017/CBO9780511801655}
}

@article{Adar2004,
  author  = {Rivka Adar and Yaakov Benenson and Gregory Linshiz and Amit Rosner and Naftali Tishby and Ehud Shapiro},
  title   = {Stochastic Computing with Biomolecular Automata},
  journal = {Proceedings of the National Academy of Sciences of the United States of America},
  volume  = {101},
  number  = {27},
  pages   = {9960--9965},
  year    = {2004},
  doi     = {10.1073/pnas.0400731101}
}

@article{Doty2012,
  author  = {David Doty},
  title   = {Theory of Algorithmic Self-Assembly},
  journal = {Communications of the ACM},
  volume  = {55},
  number  = {12},
  pages   = {78--88},
  year    = {2012},
  doi     = {10.1145/2380656.2380675}
}

@article{BarLev2025,
  author  = {Daniella Bar-Lev and Omer Sabary and Ryan Gabrys and Eitan Yaakobi},
  title   = {Cover Your Bases: How to Minimize the Sequencing Coverage in {DNA} Storage Systems},
  journal = {IEEE Transactions on Information Theory},
  volume  = {71},
  number  = {1},
  pages   = {192--218},
  year    = {2025},
  doi     = {10.1109/TIT.2024.3496587}
}

@article{Gruica2025,
  author  = {Anina Gruica and Daniella Bar-Lev and Alberto Ravagnani and Eitan Yaakobi},
  title   = {A Combinatorial Perspective on Random Access Efficiency for {DNA} Storage},
  journal = {IEEE Transactions on Information Theory},
  volume  = {71},
  number  = {12},
  pages   = {9395--9412},
  year    = {2025},
  doi     = {10.1109/TIT.2025.3623202}
}

@article{Boruchovsky2026,
  author  = {Avital Boruchovsky and Ohad Elishco and Ryan Gabrys and Anina Gruica and Itzhak Tamo and Eitan Yaakobi},
  title   = {Making It to First: The Random Access Problem in {DNA} Storage},
  journal = {IEEE Transactions on Information Theory},
  volume  = {72},
  number  = {8},
  pages   = {5623--5638},
  year    = {2026},
  doi     = {10.1109/TIT.2026.3699871}
}

@article{Bodur2026,
  author  = {{\c S}eyma Bodur and Stefano Lia and Hiram H. L{\'o}pez and Rati Ludhani and Alberto Ravagnani and Lisa Seccia},
  title   = {The Random Variables of the {DNA} Coverage Depth Problem},
  journal = {IEEE Transactions on Information Theory},
  volume  = {72},
  number  = {9},
  pages   = {6493--6503},
  year    = {2026},
  doi     = {10.1109/TIT.2026.3711141}
}

@article{GruicaMontanucciZullo2026,
  author  = {Anina Gruica and Maria Montanucci and Ferdinando Zullo},
  title   = {The Geometry of Codes for Random Access in {DNA} Storage},
  journal = {Designs, Codes and Cryptography},
  volume  = {94},
  number  = {5},
  year    = {2026},
  doi     = {10.1007/s10623-026-01857-3},
  note    = {Article 114}
}

@misc{VanDeVoordeZullo2026,
  author        = {Van de Voorde, Geertrui and Zullo, Ferdinando},
  title         = {Weak Arcs and Applications to the {DNA}-Based Storage Access Problem},
  year          = {2026},
  eprint        = {2608.19550},
  archivePrefix = {arXiv},
  howpublished  = {arXiv preprint arXiv:2608.19550},
  url           = {https://arxiv.org/abs/2608.19550}
}

@inproceedings{WangYaakobi2026,
  author        = {Chen Wang and Eitan Yaakobi},
  title         = {Random Access in {DNA} Storage: Algorithms, Constructions, and Bounds},
  booktitle     = {2026 IEEE International Symposium on Information Theory (ISIT)},
  year          = {2026},
  doi           = {10.1109/ISIT62367.2026.11653637},
  note          = {Extended version: arXiv:2601.07053},
  eprint        = {2601.07053},
  archivePrefix = {arXiv},
  url           = {https://arxiv.org/abs/2601.07053}
}

@misc{GruicaPetrilloZullo2026,
  author        = {Anina Gruica and Antonio Petrillo and Ferdinando Zullo},
  title         = {The Generalized Random Access Problem for Linear Codes},
  year          = {2026},
  eprint        = {2608.20152},
  archivePrefix = {arXiv},
  howpublished  = {arXiv preprint arXiv:2608.20152},
  url           = {https://arxiv.org/abs/2608.20152}
}

@misc{BarLevCodedRetrieval,
  author        = {Daniella Bar-Lev},
  title         = {Coded Information Retrieval for Block-Structured {DNA}-Based Data Storage},
  year          = {2026},
  eprint        = {2603.17154},
  archivePrefix = {arXiv},
  howpublished  = {arXiv preprint arXiv:2603.17154},
  url           = {https://arxiv.org/abs/2603.17154}
}

@misc{HofmeisterBitarYaakobi2026,
  author        = {Christoph Hofmeister and Rawad Bitar and Eitan Yaakobi},
  title         = {Random Access Expectation in {DNA} Data Storage: Asymptotic Analysis Through Fountain Codes},
  year          = {2026},
  howpublished  = {arXiv preprint arXiv:2605.10919},
  eprint        = {2605.10919},
  archivePrefix = {arXiv},
  note          = {Version 3},
  doi           = {10.48550/arXiv.2605.10919},
  url           = {https://arxiv.org/abs/2605.10919v3}
}

@misc{BarLevElishcoGabrys2026,
  author        = {Daniella Bar-Lev and Ohad Elishco and Ryan Gabrys},
  title         = {A {$\pi/4$} Lower Bound for Random Access in Symmetric Linear Ensembles under Optimal Decoding},
  year          = {2026},
  howpublished  = {arXiv preprint arXiv:2610.38446v1},
  eprint        = {2610.38446},
  archivePrefix = {arXiv},
  url           = {https://arxiv.org/abs/2610.38446v1}
}

@article{SabaryKiahSiegelYaakobi2024,
  author  = {Omer Sabary and Han Mao Kiah and Paul H. Siegel and Eitan Yaakobi},
  title   = {Survey for a Decade of Coding for {DNA} Storage},
  journal = {IEEE Transactions on Molecular, Biological, and Multi-Scale Communications},
  volume  = {10},
  number  = {2},
  pages   = {253--271},
  year    = {2024},
  doi     = {10.1109/TMBMC.2024.3403488}
}

\end{document}